\documentclass[prx,letterpaper,aps,10pt,superscriptaddress,twocolumn,floatfix,showpacs]{revtex4-1}
\usepackage{graphicx}
\usepackage{amsmath}
\usepackage{amsfonts}
\usepackage{float}
\usepackage{amssymb}
\usepackage{epsfig}
\usepackage{epstopdf}
\DeclareGraphicsExtensions{.pdf,.eps,.png,.jpg,.mps}
\usepackage[pdftex]{color}
\usepackage{amsmath,graphicx,amssymb,braket,xcolor,subfigure,upgreek}
\usepackage[colorlinks, linkcolor=blue, citecolor=blue, urlcolor=blue, breaklinks=true]{hyperref}
\usepackage{microtype}
\usepackage{bbm}
\usepackage{color}
\usepackage{dsfont}
\usepackage[english]{babel}
\usepackage[utf8]{inputenc}
\usepackage{textcomp}

\newcommand{\C}[1]{\textcolor{black}{#1}}

\DeclareMathOperator{\sinc}{sinc}
\DeclareMathOperator{\erf}{erf}

\begin{document}

\title{Molecular polaritonics in dense mesoscopic disordered ensembles}
\author{C.~Sommer}
\affiliation{Max Planck Institute for the Science of Light, Staudtstra{\ss}e 2,
D-91058 Erlangen, Germany}
\author{M.~Reitz}
\affiliation{Max Planck Institute for the Science of Light, Staudtstra{\ss}e 2,
D-91058 Erlangen, Germany}
\affiliation{Department of Physics, University of Erlangen-Nuremberg, Staudtstra{\ss}e 2,
D-91058 Erlangen, Germany}
\author{F.~Mineo}
\affiliation{Max Planck Institute for the Science of Light, Staudtstra{\ss}e 2,
D-91058 Erlangen, Germany}
\affiliation{Department of Physics, University of Erlangen-Nuremberg, Staudtstra{\ss}e 2,
D-91058 Erlangen, Germany}
\author{C.~Genes}
\affiliation{Max Planck Institute for the Science of Light, Staudtstra{\ss}e 2,
D-91058 Erlangen, Germany}
\affiliation{Department of Physics, University of Erlangen-Nuremberg, Staudtstra{\ss}e 2,
D-91058 Erlangen, Germany}
\date{\today}

\begin{abstract}
\C{We study the dependence of the vacuum Rabi splitting (VRS) on frequency disorder, vibrations, near-field effects and density in molecular polaritonics. In the mesoscopic limit, static frequency disorder alone can already introduce a loss mechanism from polaritonic states into a dark state reservoir, which we quantitatively describe, providing an analytical scaling of the VRS with the level of disorder. Disorder additionally can split a molecular ensemble into donor-type and acceptor-type molecules and the combination of vibronic coupling, dipole-dipole interactions and vibrational relaxation induces an incoherent FRET (F\"{o}rster resonance energy transfer) migration of excitations within the collective molecular state. This is equivalent to a dissipative disorder and has the effect of saturating and even reducing the VRS in the mesoscopic, high-density limit. Overall, this analysis allows to quantify the crucial role played by dark states in cavity quantum electrodynamics with mesoscopic, disordered ensembles.}
\end{abstract}

\pacs{42.50.Ar, 42.50.Lc, 42.72.-g}

\maketitle

\section{Introduction}
The strength of light-matter coherent exchanges is enhanced when confined light modes, such as provided by optical cavities, are utilized. For $\mathcal{N}$ ideal two-level quantum emitters, equally coupled to a cavity mode, a collective enhancement proportional to $\mathcal{N}^{1/2}$ can be obtained~\cite{tavis1968exact}. This is evident in the scaling of the collective vacuum Rabi splitting (VRS) in cavity quantum electrodynamics (cQED)~\cite{haroche1989cavity,berman1994cavity,walther2006cavity}. In the particular case where more complex emitters, such as \C{organic} molecules \C{(J-Aggregates, dye molecules, etc.)}, are collectively coupled to optical or plasmonic resonators, these standard results of cQED have been extensively invoked to describe the collective Rayleigh scattering loss from a cavity~\cite{golombek2020collective}, the modification of energy transfer and transport~\cite{schachenmayer2015cavity,feist2015extraordinary,zhong2016non,zhong2017energy,feist2017long,reitz2018energy}, charge transport~\cite{orgiu2015conductivity,hagenmuller2017cavity,hagenmuller2018cavity,zeb2020incoherent} or chemical reactions in the presence of strong light-matter interactions~\cite{galego2016suppressing,herrera2016cavity,herrera2020molecular,zhou2020polariton}. However, molecular polaritonics is characterized by emitters with large inhomogenous broadening, coupled to local vibrational baths and with strong near-field interactions, in which case analytical approaches are typically limited to only a few molecules and often with only one vibrational mode~\cite{herrera2017dark,neuman2018origin,zeb2018exact,reitz2019langevin}. The numerical complexity of treating many electronic and vibrational degrees of freedom renders such problems hard to solve even with extensive simulations~\cite{pino2018tensor,groenhof2019tracking,mordovina2020polaritonic}.\\
\begin{figure}[t]
\includegraphics[width=0.85\columnwidth]{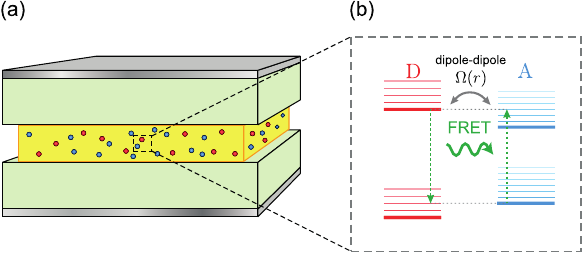}
\caption {a) Cavity-enclosed dense, disordered molecular ensemble, with electronic transitions subject to vibronic and near-field couplings is naturally split (owing to frequency disorder) between donor-like (red) and acceptor-like (blue) molecules. b) Schematics of processes leading to the incoherent FRET migration of excitations. The near-field coupling (with distance dependent strength $\Omega(r)$), followed by rapid vibrational relaxation within the vibrational manifold of the excited electronic acceptor state, leads to a unidirectional flow of energy.}
\label{fig1}
\end{figure}
\indent \C{We propose here a fully analytical approach which allows to quantify the effect of disorder on light-matter interactions in the strong coupling regime. In a first step, we introduce the formalism for the case of pure two-level systems (involving electronic transitions only) with general applicability to cQED with atoms, quantum dots, superconducting qubits, etc.~\cite{munstermann2000observation, fink2009dressed, kubo2010strong}. In a second step, we exemplify the application of this formalism to more complex systems involving electron-phonon interactions and in particular to molecular polaritonics~\cite{herrera2020molecular}.\\
\indent The two main conceptual ingredients of our approach consist in the move to a collective basis for $\mathcal{N}$ emitters and the occurrence of a natural averaging in the mesoscopic limit (as opposed to averaging over many realizations ~\cite{houdre1996vacuum,diniz2011strongly,kurucz2011spectroscopic,debnath2019collective}). As widely acknowledged, polaritons are formed by one \textit{bright} superposition state hybridized with light while the rest of $\mathcal{N}-1$ \textit{dark} states are only indirectly coupled owing to disorder~\cite{agranovich1998biexcitons, litinskaya2004fast,ballestero2016uncoupled,herrera2017dark,ribeiro2018polariton,botzung2020dark}. We take an open system dynamics approach to derive an analytical rate for the irreversible loss of energy from polaritonic states into the dark state manifold accompanied by a degradation of the VRS. In the bare basis, we elucidate the reduction of the VRS by showing that particles which are too far detuned or too lossy can fall out of the macroscopic polaritonic superposition.\\
\indent We then apply our formalism to molecular polaritonics where the interplay between static disorder, near-field couplings and vibrational relaxation leads to a FRET process characterized by incoherent transfer of excitations from energetically higher donor-type to lower frequency, acceptor-type molecules (see Fig.~\ref{fig1}). We map this problem into an incoherent dynamics in Lindblad form describing migration of excitation at rates analytically computable and derive the scaling law for the VRS with density applying the open system dynamics previously derived for pure two-level systems.}\\
\indent The paper is structured as follows: we introduce the Tavis-Holstein-Cummings model for $\mathcal{N}$ molecules each with two electronic and $n$ vibronic degrees of freedom coupled to a confined cavity mode in Sec.~\ref{Sec2}. We then proceed by analyzing the cavity transmission in the presence of frequency disorder in Sec.~\ref{Sec3} and show how a mesoscopic average leads to a decay of polaritons into the dark state manifold. To characterize the degree of participation of quantum emitters to the collective strong coupling condition, we introduce a measure of macroscopicity of quantum  superpositions reaching value $\mathcal{N}$ for perfect superpositions and unity for complete mixtures. The effects of dipole-dipole couplings together with vibrational relaxation are taken into account in Sec.~\ref{Sec4} and the elimination of the dark state reservoir is revisited, this time including the process of incoherent excitation migration within the molecular ensemble. Finally, we conclude and present an outlook in Sec.~\ref{Sec5}.

\section{Model}
\label{Sec2}
We consider $\mathcal{N}$ molecules indexed by $j=1,...\mathcal{N}$ with electronic states $\ket{g}_j$ and $\ket{e}_j$ (lowering operator $\sigma_j=\ket{g}_j\bra{e}_j$) separated by energy splittings $\omega_j$ ($\hbar=1$) inhomogeneously distributed around $\omega$ with a distribution function $p(\delta)$ normalized to unity $\int_{-\infty}^{\infty}p(\delta)d\delta=1$. In particular we choose $p(\delta) = (1/\sqrt{2\pi w^2})e^{-\delta^2/(2w^2)}$. We write each molecule frequency splitting as $\omega+\delta_j$ where the average around the central frequency vanishes $\braket{\delta_j}_\text{cl}=0$ while the variance is $\braket{\delta_j^2}_\text{cl}=w^2$. The molecules are randomly spatially distributed within a volume $\mathcal{V}$ at positions $\mathbf{r}_j$. Each molecule exhibits a number $n$ of nuclear coordinates each with frequency $\nu_k$ (with $k=1...n$) with harmonic motion described by the annihilation operators $b_{jk}$ such that $\left[b_{jk},b_{jk}^\dagger\right]=1$. The vibronic couplings are modeled as Holstein terms with Huang-Rhys factors $\lambda_k^2$ stemming from a difference in the equilibrium positions of the ground and excited electronic potential landscapes.\\
\indent For high densities, the near-field dipole-dipole interactions at rates $\Omega_{jj'}$ are dependent on the separation (with a standard $|\mathbf{r}_j-\mathbf{r}_{j'}|^{-3}$ dependence) and relative orientation of transition dipoles. The dipole-dipole Hamiltonian is
\begin{equation}
\mathcal{H}_\text{d-d}=\textstyle \sum_{j\neq j'}\Omega_{jj'}\sigma^\dagger_j\sigma_{j'}
\end{equation}
and describes an excitation transfer via a virtual photon exchange.
The free Hamiltonian is (see Ref.~\cite{reitz2019langevin})
\begin{equation}
\mathcal{H}_0=\sum_{j=1}^{\mathcal{N}}\left[\omega+\delta_j+\sum_{k=1}^{n}\lambda_k^2 \nu_k\right]\sigma^\dagger_j\sigma_j+\sum_{j=1}^{\mathcal{N}}\sum_{k=1}^{n}\nu_k b_{jk}^\dagger b_{jk},
\end{equation}
and adds to the vibronic coupling Hamiltonian~\cite{holstein1959study}
\begin{equation}
\mathcal{H}_{\text{Hol}}=-\sum_{j=1}^{\mathcal{N}}\sum_{k=1}^{n}\lambda_k \nu_k \sigma_j^\dagger\sigma_j (b_{jk}^\dagger+b_{jk}).
\end{equation}
The vibronic coupling is obtained as a harmonic approximation of a Morse potential surface by expanding the electronic potential landscapes around their minima: the difference between the minima in the ground and excited state leads then to the Huang-Rhys factors $\lambda_k^2$. Such a model is widely employed \cite{herrera2017dark, neuman2018origin, wu2016when, kansanen2019theory} especially for molecules in condensed matter environments, as fast vibrational relaxation insures that states with more than one vibrational excitation are never reached.\\
\indent The cavity mode is described by bosonic operator $a$ at frequency $\omega_c$ coupled with $g_j(\mathbf{r}_j)\equiv g_j$ to each molecule. The Tavis-Cummings Hamiltonian is then
\begin{equation}
\mathcal{H}_\text{TC}= a \sum_{j=1}^{\mathcal{N}} g_j \sigma^\dagger_j+ a^\dagger \sum_{j=1}^{\mathcal{N}} g_j^* \sigma_j.
\end{equation}
This is a simplification of the Dicke model when neglecting counter-rotating terms such as $a^\dagger \sigma_j^\dagger$. While some current experiments operate on the brink of the ultrastrong coupling regime~\cite{lidzey1999room, schwartz2011reversible, held2018ultrastrong}, polariton dynamics is well reproduced within this approximation.\\
\indent We then proceed by writing the master equation of the system
\begin{equation}
\partial_t \rho=i[\rho, \mathcal{H}]+\mathcal{L}[\rho],
\end{equation}
where the dissipative dynamics is included in the Lindblad part. For a collapse operator $\cal{O}$ with rate $\gamma_{\cal{O}}$ the Lindblad term applied to a density operator $\rho$ is
\begin{equation}
{\cal{L}}_{\cal{O}}[\rho]=\gamma_{\cal{O}}\left\{2\cal{O}\rho \cal{O}^\dagger-\rho\cal{O}^\dagger\cal{O}-\cal{O}^\dagger\cal{O}\rho\right\}.
\end{equation}
All channels of dissipation are then modeled as standard Lindblad superoperators with collapse operators $a,\sigma_j,b_{jk}$ and loss rates $\kappa,\gamma,\Gamma_k$.


\section{Effects of disorder}
\label{Sec3}
We will first show the effect of frequency disorder as the occurrence of dark state resonances in the (linear) spectral response of the cavity when driven with an external (weak) laser source. The pump is modelled via the following Hamiltonian
\begin{equation}
\mathcal{H}_d=i\eta(a^\dagger e^{-i\omega_\ell t}-a e^{i\omega_\ell t}),
\end{equation}
where the pump frequency is $\omega_\ell$ and the weak drive amplitude is $\eta$. The equations of motion for the averages $\alpha=\braket{a}$ and $\beta_j=\braket{\sigma_j}$ then read
\begin{subequations}
\label{eqsmotion}
\begin{align}
\dot{\beta}_j &= -i(\omega - \omega_\ell + \delta_j -i\gamma)\beta_j -ig_j\alpha, \\
\dot{\alpha} &= -i(\omega_{c}-\omega_\ell -i\kappa)\alpha - i\textstyle{\sum}_j g^{*}_j\beta_j + \eta,
\end{align}
\end{subequations}
In a more compact form we can write
\begin{equation}
\dot{\mathbf{v}} = - i M\mathbf{v} + \mathbf{v}_d,
\end{equation}
where the vector of amplitudes is $\mathbf{v} = (\beta_1, \dots ,\beta_\mathcal{N},\alpha)^{\top}$, driving is included also in vector form as $\mathbf{v}_d = (0, \dots , 0,\eta)^{\top}$ and the drift matrix is explicitly given in Appendix \ref{A2}.

\subsection{Steady state cavity transmission}

\begin{figure*}[t]
\includegraphics[width=1.95\columnwidth]{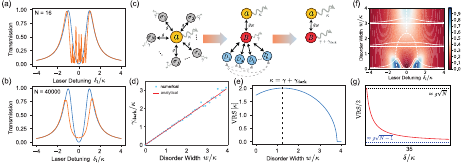}
\caption{a) Transmission for $\mathcal{N}=16$, $g = 0.25$, $\gamma=10^{-2}$ (units of $\kappa$) for $w=0$ and $w = 0.5$. (b) In the mesoscopic limit ($\mathcal{N}=4\times10^4$) with $g = 0.005$ the dark state peaks are smoothed out. Loss into dark states leads to a reduction of polariton height and increase of splitting (orange - $w=0.5$) compared to the $w=0$ case (blue). (c) Elimination of the dark state reservoir. The transformation to a collective basis sees the cavity mode $a$ solely coupled to a bright mode at rate $g_\mathcal{N}$. The dark state manifold provides a loss channel at rate $\gamma_\text{dark}$. (d) Decay rate $\gamma_{\text{dark}}$ as a function of $w$. (e) VRS with increasing disorder $w$. (f) Exact numerical results for the cavity transmission as a function of the width $w$ for $\mathcal{N}=4\times10^4$. Dashed line shows the maxima of the transmission under the Markovian approximation. (g) \C{VRS degradation from $g\sqrt{\mathcal{N}}$ to $g\sqrt{\mathcal{N}-1}$ as a particle is removed from the polaritonic superposition by increasing its detuning from the cavity resonance}.}
\label{fig2}
\end{figure*}

In steady state, the equations above lead to the normalized cavity amplitude transmission $t=\kappa \braket{a}/\eta$ expressed as
\begin{align}
t=\kappa\left[\kappa+i (\omega_c-\omega_\ell)+\sum_{j=1}^{\mathcal{N}}\frac{|g_j|^2}{\gamma+i (\omega-\omega_\ell)+i\delta_j}\right]^{-1},
\end{align}
valid also for positioning and orientational disorder with randomized couplings $g_j$. The effect of orientational disorder is a trivial renormalization of the collective coupling from $g\sqrt{\mathcal{N}}$ to $g\sqrt{\mathcal{N}/2}$ (for a completely random orientation of the molecular dipoles, as discussed in Appendix \ref{A5}).\\
\indent In the following we restrict the discussion to the case of identical couplings $g_{j} = g$ (for all $j$). For a given realization of disorder, Fig.~\ref{fig2}a shows two polaritonic peaks at $\pm g\sqrt{\mathcal{N}}$ obtained by the hybridization of a symmetric collective state to the cavity field. Non-zero disorder introduces couplings to $\mathcal{N}-1$ orthogonal asymmetric states visible in the cavity transmission as unequal height peaks between the polaritons. In the mesoscopic limit (see Fig.~\ref{fig2}b), a natural averaging occurs which leads to a smoothing out of the additional peaks. Also, the polariton's height is decreased while their width is increased suggesting a loss mechanism which we will quantitatively address in the following in a transformed bright-dark basis.
\subsection{Bright-dark state dynamics}

\indent We start with $w=0$ and note that the cavity couples only to a symmetric superposition $\hat{B}=\textstyle \sum_j \sigma_j/\sqrt{\mathcal{N}}$, i.e.~a \textit{bright state}, with a collective coupling strength $g_\mathcal{N} = \sqrt{\mathcal{N}}g$. The other $\mathcal{N}-1$ combinations define \textit{dark states} which are generally obtainable by a Gram-Schmidt algorithm that leads to all vectors orthogonal to the bright state one and to each other. However, for the simplest case $g_j=g$ a straightforward choice of coefficients is indicated by a discrete Fourier transform
\begin{equation}
\hat{D}_{k} = \frac{1}{\sqrt{\mathcal{N}}} \sum^{\mathcal{N}}_{j = 1}e^{-i2\pi jk/\mathcal{N}}\sigma_j.
\end{equation}
We index the dark state manifold for $k=1,\dots,\mathcal{N}-1$ and note that for $k=\mathcal{N}$ we recover the bright state $\hat{D}_{\mathcal{N}} = \hat{B}$. The equations of motion for averages $\mathcal{B}=\braket{\hat{B}}$, $\mathcal{D}=\braket{\hat{D}}$ and $\alpha$ become (in a frame rotating at the central emitter frequency $\omega$)
\begin{subequations}
\label{Eqs.Dk}
\begin{align}
\dot{\mathcal{D}}_k &= -\gamma \mathcal{D}_k-i\sum^{\mathcal{N}}_{k' = 1}\Delta_{kk'} \mathcal{D}_{k'} -i g_\mathcal{N}\alpha \delta_{k\mathcal{N}},\\
\dot{\alpha} &= -i(\delta-i\kappa)\alpha-ig^{*}_\mathcal{N} \mathcal{D}_\mathcal{N}+\eta,
\end{align}
\end{subequations}
with $\delta=\omega_c-\omega_\ell$ and the couplings are defined as Fourier transforms of the disorder distribution
\begin{equation}
\Delta_{kk'} = \frac{1}{\mathcal{N}}\sum^{\mathcal{N}}_{j = 1} \delta_j e^{-i 2\pi j(k-k')/\mathcal{N}}.
\end{equation}
For $k=\mathcal{N}$, the equations above indicate that the bright state is the only one coupled to the cavity mode with the expected collective rate $g_\mathcal{N}$. However, disorder induces couplings to the whole manifold of dark states and within the dark manifold as well.\\
\indent We can more compactly write the equations above as $\partial_t\mathbf{V}=-iM_\text{coll} \mathbf{V}$ where $\mathbf{V}=(\mathcal{D}_1,\ldots,\mathcal{D}_{\mathcal{N}-1}, \mathcal{B}, \alpha)^\top$ and the drift matrix is
\begin{equation}
  M_\text{coll} =\left(
    \begin{array}{ccccc}
      (\bar{\delta}-i\gamma) & \Delta_{12} & \ldots & \Delta_{1\mathcal{N}} & 0\\
      \Delta_{21} & (\bar{\delta}-i\gamma) & \ldots & \Delta_{2\mathcal{N}} & 0 \\
      \vdots & \vdots & \ddots & \vdots & \vdots\\
      \Delta_{\mathcal{N}1} & \Delta_{\mathcal{N}2} & \ldots & (\bar{\delta}-i\gamma) & g_\mathcal{N}\\
			0 & 0 & \ldots & g^{*}_\mathcal{N}  & (\delta -i\kappa) \\
    \end{array}
  \right),
\end{equation}
where the average of the disorder distribution (expected to vanish in the mesoscopic limit) is $\bar{\delta}=\Delta_{kk}=\textstyle\sum^{\mathcal{N}}_{j = 1} \delta_j/\mathcal{N}$. This matrix can be put in a more convenient form due to the structure of the terms $\Delta_{l\mathcal{N}}$. Considering the relations $\Delta_{\mathcal{N}l} = \Delta_{\mathcal{N}-l\mathcal{N}}$ and $\Delta_{1\mathcal{N}} = \Delta_{\mathcal{N}\mathcal{N}-1}$, we can rewrite the matrix as
\begin{equation}
M_\text{coll} = \left(
\begin{array}{cc}
      M_\text{red} & \mathbf{p} \\
			\mathbf{p}^{\dagger} & (\delta -i\kappa)
    \end{array}
\right),
\end{equation}
with $\mathbf{p} =  (0,\dots, 0 , g_\mathcal{N})^{\top}$. Notice that here we have separated between the cavity and matter states by introducing the reduced matrix of dimensions $\mathcal{N} \times \mathcal{N}$ (referring to the matter part) which assumes the following form
\begin{equation}
M_\text{red} = \left(
\begin{array}{cccccc}
      (\bar{\delta}-i\gamma) & \Delta_{\mathcal{N}-1\mathcal{N}} & \Delta_{\mathcal{N}-2\mathcal{N}} & \ldots & \Delta_{1\mathcal{N}} \\
      \Delta_{1\mathcal{N}} & (\bar{\delta}-i\gamma) & \Delta_{\mathcal{N}-1\mathcal{N}} & \ldots & \Delta_{2\mathcal{N}}  \\
			\Delta_{2\mathcal{N}} & \Delta_{1\mathcal{N}} & ( \bar{\delta}-i\gamma) & \ldots & \Delta_{3\mathcal{N}}  \\
      \vdots & \vdots & \vdots & \ddots & \vdots\\
      \Delta_{\mathcal{N}-1\mathcal{N}} & \Delta_{\mathcal{N}-2\mathcal{N}} & \Delta_{\mathcal{N}-3\mathcal{N}} & \ldots & (\bar{\delta}-i\gamma)
    \end{array}
\right).
\end{equation}
The eigenvectors of the cyclic matrix $M_\text{red}$ are given by
$v_{j} = (1/\sqrt{\mathcal{N}})(1, \xi_{j}, \xi^{2}_{j}, \dots, \xi^{\mathcal{N}-1}_{j})^{\top}$ for $j \in \{1,\dots, \mathcal{N}-1\}$ where $\xi_{j} = \exp(i2\pi j/\mathcal{N})$. The eigenvalues are given by $\lambda_j = (\bar{\delta}-i\gamma) + \Delta_{\mathcal{N}-1\mathcal{N}} \xi_j + \dots + \Delta_{1\mathcal{N}} \xi^{\mathcal{N}-1}_j$.\\

\subsection{Elimination of the dark reservoir}
The procedure we will employ roughly follows the illustration in Fig.~\ref{fig2}c showing first the identification of a bright state and then the elimination of the dark reservoir resulting in an effective unidirectional loss of energy from the polaritonic states. The elimination of the dark state manifold can be done in an exact way without making a Markovian approximation, which would imply that the dark state reservoir has no memory and therefore it would allow to set all derivatives of $\mathcal{D}_k$ to zero. Instead, we formally integrate the equations for $\mathcal{D}_k$ to obtain (Appendix \ref{A4})
\begin{equation}
\dot{\mathcal{B}}(t) = -(\gamma +i \bar{\delta})\mathcal{B}(t)-\int_{-\infty}^{\infty}dt' f(t-t')\mathcal{B}(t') -i g_\mathcal{N}\alpha,
\end{equation}
and obtain a memory kernel describing a generally non-Markovian loss process. In the mesoscopic limit, one finds $f(t-t')\approx\Theta(t-t')w^2e^{-i(\bar{\delta}-i\gamma)(t-t')} \sinc(2w(t-t'))$. Now one can identify a Markovian limit as a particular case of wide frequency distributions $w\gg\gamma$. The Markovian result is then simply reproduced by seeing that the kernel $f(t-t')$ naturally tends to a delta function. In such a case the treatment can be simplified by setting all derivatives to zero in Eqs.~\eqref{Eqs.Dk} to find the dark state amplitudes
\begin{equation}
\mathcal{D}_k=-\sum_{k'}({\mathcal{M}^{-1}})_{kk'}\Delta_{k'\mathcal{N}} \mathcal{B}.
\end{equation}
The matrix $\mathcal{M}$ has dimensions $(\mathcal{N}-1)\times (\mathcal{N}-1)$  and represents the part of the matrix $M_\text{red}$ referring to the dark states only
\begin{equation}
  \mathcal{M} =\left(
    \begin{array}{cccc}
      (\bar{\delta}-i\gamma) & \Delta_{12} & \ldots & \Delta_{1(\mathcal{N}-1)} \\
      \Delta_{21} & (\bar{\delta}-i\gamma) & \ldots & \Delta_{2(\mathcal{N}-1)}  \\
      \vdots & \vdots & \ddots & \vdots \\
      \Delta_{(\mathcal{N}-1)1} & \Delta_{(\mathcal{N}-1)2} & \ldots & (\bar{\delta}-i\gamma) \\
    \end{array}
  \right).
\end{equation}
Replacing the eliminated variables into the equation of motion for the bright mode we obtain the effective dissipative dynamics
\begin{equation}
\dot{\mathcal{B}} = -i\left[ (\bar{\delta}-\delta_\text{dark}) -i(\gamma+\gamma_\text{dark})\right]\mathcal{B}-i g_\mathcal{N}\alpha,
\end{equation}
where the effect of the reservoir is to induce an effective frequency shift $\delta_\text{dark}$ and loss rate $\gamma_\text{dark}$ obtained as the real and imaginary parts, respectively, of the following expression
\begin{align}
\label{DrkRes.1}
\delta_\text{dark}&+i \gamma_\text{dark}=\sum_{k,k'=1}^{\mathcal{N}-1}\Delta_{\mathcal{N}k}(\mathcal{M}^{-1})_{kk'}\Delta_{k'\mathcal{N}}\\
&=\frac{1}{\mathcal{N}^2}\sum_{j,j'=1}^{\mathcal{N}} \sum_{k,k'=1}^{\mathcal{N}-1} \delta_j\delta_{j'}(\mathcal{M}^{-1})_{kk'} e^{-i 2\pi (j k- j' k') /\mathcal{N}}.\nonumber
\end{align}

In the mesoscopic limit of large $\mathcal{N}$, one can further simplify the expression of the loss rate to find extremely simple scaling laws of the decay rate induced by the dark state manifold (see Appendix \ref{A3} for details)
\begin{equation}
\gamma_\text{dark} = \left\{ \begin{array}{ccc} w^2/\gamma & \qquad \text{for} \qquad & \gamma \gg w \\  \pi w/4 & \qquad \text{for} \qquad & w \gg \gamma \end{array}. \right.
\end{equation}
The analytical results are in excellent agreement with numerical simulations (see Fig.~\ref{fig2}d) which also indicate that both $\bar{\delta}$ and $\delta_\text{dark}$ vanish. From here we can deduce the dependence of the VRS on disorder which can be obtained by diagonalizing the dynamics in the reduced cavity-bright state subspace to lead to
\begin{eqnarray}
\label{theoryvrs}
\text{VRS} \approx \Im \left\{2\sqrt{\left(\gamma+\gamma_\text{dark}-\kappa\right)^2/4-g_\mathcal{N}^2}\right\}.
\end{eqnarray}
This is an important result later generalized to molecules to analytically quantify the effect of time-dependent disorder associated with continuous shifting of electronic resonances via vibronic driving. The VRS is illustrated in Fig.~\ref{fig2}e as a function of increasing disorder. The
initial increase of the polariton splitting occurs in the case $\kappa>\gamma$ as the maximum value of $2g_\mathcal{N}$ is reached when $\gamma+\gamma_\text{dark}=\kappa$. The complete degradation of the strong coupling condition occurs when the disorder level is of the order of the cavity photon loss. The validity of the Markovian approximation is graphically illustrated in Fig.~\ref{fig2}f. In Appendix \ref{A4}, we perform a more in-depth analysis of the non-Markovian regime by means of the quantum Langevin equations approach.  \\

\subsection{Reduction of VRS owing to far detuned particles}
Let us now provide further clarifications of the mechanism of VRS degradation owed to inhomogeneous broadening by reverting our analysis to the alternative bare basis approach. To this end we start with the standard scenario of $\mathcal{N}$ identical particles coupled equally to the cavity field
\begin{subequations}
\begin{align}
\dot{\beta}_j &= -\gamma\beta_j - ig\alpha,\\
\dot{\alpha} &= -\kappa\alpha -ig\sum^\mathcal{N}_{j=1}\beta_j.
\end{align}
\end{subequations}
and notice immediately that the equations can be described in terms of a single bright state such that
\begin{subequations}
\begin{align}
\dot{\mathcal{B}}_\mathcal{N} &= -\gamma\mathcal{B}_\mathcal{N} - ig\sqrt{\mathcal{N}}\alpha, \\
\dot{\alpha} &= -\kappa\alpha -ig\beta_1 -i g\sqrt{\mathcal{N}}\mathcal{B}_\mathcal{N}.
\end{align}
\end{subequations}
This indicates that the problem is simply described by a single collective mode strongly coupled at $g\sqrt{\mathcal{N}}$ coupling while the dark manifold is completely decoupled and does not play any role in the dynamics. Now instead we assume $\mathcal{N} -1$ particles identically and resonantly coupled to the cavity mode while an additional particle is detuned by $\delta$. We can rewrite the equations above now in terms of the $\mathcal{N}-1$ bright state
\begin{subequations}
\begin{align}
\dot{\beta}_1 &= -\gamma \beta_1-i\delta \beta_1 - ig\alpha, \\
\dot{\mathcal{B}}_{\mathcal{N}-1} &= -\gamma\mathcal{B}_{\mathcal{N} -1} - ig\sqrt{\mathcal{N} -1}\alpha, \\
\dot{\alpha} &= -\kappa\alpha -ig\beta_1 -i\sqrt{\mathcal{N} -1}g\mathcal{B}_{\mathcal{N} -1}.
\end{align}
\end{subequations}
The bright state is similarly defined as: $\mathcal{B}_{\mathcal{N}-1} = (1/\sqrt{\mathcal{N}-1})\sum^\mathcal{N}_{j=2}\beta_j$. A close inspection of the above equations shows that when $\delta$ is detuned from the cavity resonance, the corresponding VRS shows a drop from $g\sqrt{\mathcal{N}}$ for $\delta=0$ to $g\sqrt{\mathcal{N} -1}$ for $\delta\gg \kappa$ (see Fig.~\ref{fig2}g). For large $w$, this behavior indicates that, when disorder is strong more particles are likely to have frequencies very far from the cavity resonance, which finally leads to the degradation of the strong coupling condition as clearly illustrated in Fig.~\ref{fig2}e.

\subsection{Macroscopicity of mesoscopic quantum superposition states}

\begin{figure*}[t]
\includegraphics[width=1.95\columnwidth]{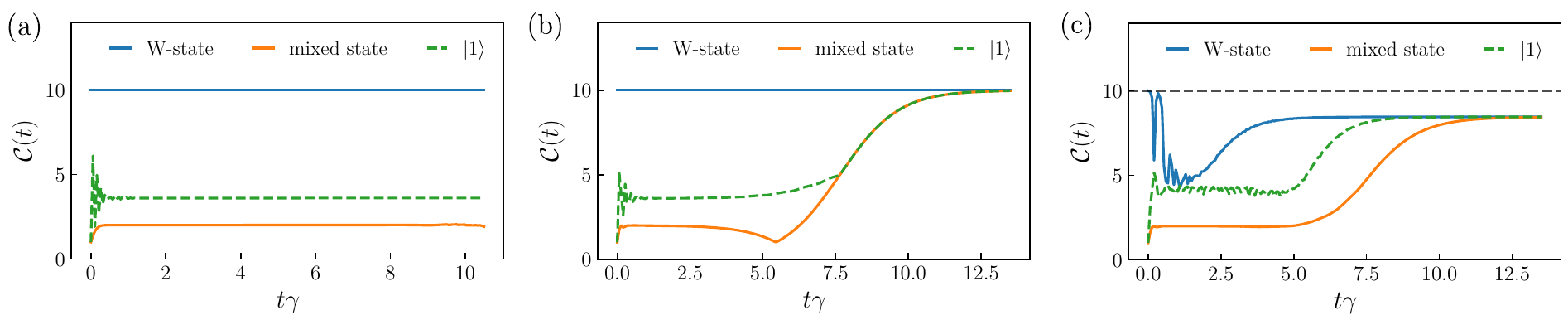}
\caption {\textit{Time dynamics of macroscopicity.} Time dynamics of macroscopicity $\mathcal{C}(t)$ for $\mathcal{N}=10$ particles for three different initial collective states of the quantum emitter ensemble: perfect superposition $\ket{W}$, completely mixed state and single particle excitation (e.g.,~$\ket{1}$). a) Undriven cavity shows conservation of macroscopicity. b) Driven cavity without disorder and $\eta=\gamma$ shows production of maximal macroscopicity in all cases. c) Driven cavity with disorder at the level $w=30\gamma$. Other parameters are: $g=40\gamma$, $\kappa=20\gamma$. Partial reduction of macroscopicity is obtained as an effect of disorder.}
\label{fig4}
\end{figure*}
As the VRS can be derived from a Hamiltonian formulation restricted to the single excitation subspace it is interesting to also investigate the connection between VRS, as observed for example in the cavity transmission and the properties of the quantum superposition state. To this end, we propose a simple measure of quantum macroscopicity
\begin{equation}
\mathcal{C}=\sum_{j,j',j\neq j'}|\braket{\sigma_j^\dagger \sigma_{j'}}_\text{red}|+1,
\end{equation}
that aims at describing the number of quantum emitters actively and identically participating in an extended quantum state. The measure is restricted to the single excitation subspace spanned by states where one individual emitter is excited by $\ket{j} = \ket{g,\dots, e_{j}, \dots, g, 0}$ for $j \in \{1,\dots, \mathcal{N}\}$. The cavity mode excitation is represented by $\ket{\mathcal{N}+1} = \ket{g,\dots,g,1}$ and the ground state is $\ket{0} = \ket{g,\dots,g,0}$. The averaging in the reduced subspace is performed with the reduced density operator
\begin{equation}
\rho_{\text{red}}=\frac{\mathcal{P}_{\mathcal{N}+1} \mathcal{P}_0 \rho \mathcal{P}_0 \mathcal{P}_{\mathcal{N}+1}}{\text{Tr}[\mathcal{P}_{\mathcal{N}+1} \mathcal{P}_0 \rho \mathcal{P}_0 \mathcal{P}_{\mathcal{N}+1}]}.
\end{equation}
The projectors $\mathcal{P}_0=\mathds{1}_{\mathcal{N}+2}-\ket{0}\bra{0}$, $\mathcal{P}_{\mathcal{N}+1}=\mathds{1}_{\mathcal{N}+2}-\ket{\mathcal{N}+1}\bra{\mathcal{N}+1}$ simply eliminate the parts of the density matrix containing the ground and the photonic state.\\
Notice that
\begin{equation}
\text{Tr}(\mathcal{P}_{\mathcal{N}+1}\mathcal{P}_0\rho \mathcal{P}_0 \mathcal{P}_{\mathcal{N}+1})=\sum_{j=1}^{\mathcal{N}} \rho_{jj},
\end{equation}
can be computed simply as $1-\rho_{00}-\rho_{\mathcal{N}+1,\mathcal{N}+1}$. This allows us to rewrite the macroscopicity with respect to components of the original density matrix $\rho$ as
\begin{equation}
\mathcal{C}(t)=\sum_{\begin{smallmatrix} j,j'=1 \\ j\neq j' \end{smallmatrix}}^{\mathcal{N}} \left|\frac{\rho_{jj'}(t)}{1-\rho_{00}(t)-\rho_{\mathcal{N}+1,\mathcal{N}+1}(t)}\right| +1.
\end{equation}

Notice that the measure is tuned such that it equals $\mathcal{\mathcal{N}}$ for a perfect W-state $\ket{W}=\textstyle \sum_j \ket{j}/\sqrt{\mathcal{N}}$ and it drops to unity for completely mixed states such as described by a density operator $\rho=\textstyle \sum_j\ket{j}\bra{j}/\mathcal{N}$. We illustrate the time dynamics of the introduced measure of macroscopicity in three distinct cases for three different initial states. We distinguish between an initial state with maximal macroscopicity (the W-state) and two states with single particle participation (mixed state versus single excitation state). The important result illustrated in Fig.~\ref{fig4}b shows that by coherently driving the cavity mode one can create maximal macroscopicity independently on the initial state. Also, in the presence of disorder, the macroscopicity is diminished for all initial states by the same amount. According to the interpretation obtained in the VRS case, one can see that for this given realization of disorder two far-detuned particles fall out of the macroscopic superposition thus diminishing $\mathcal{C}$ by 2 [Fig.~\ref{fig4}c]. In conclusion, we find that cavity driving can create macroscopicity while disorder and strong donor behavior can destroy it. It will therefore be interesting to extend such a measure beyond the single excitation subspace and to pursue in the future an analysis of the connection between collective strong coupling and the macroscopicity of quantum superposition states.

\section{Reduction of the VRS in molecular polaritonics}
\label{Sec4}

Light-matter interactions in molecular ensembles are strongly modified in the presence of electron-vibron coupling as well as by the incoherent dynamics of molecular vibrations. In addition, in standard experimental situations densities are very high meaning that near field effects such as dipole-dipole couplings can play an important role. We will provide here a semi-analytical approach incorporating the competition between static disorder, vibronic and dipole-dipole couplings together with  vibrational relaxation which can lead to a migration of excitation from a higher energy molecule (donor) to a lower energy one (acceptor) (as illustrated in Fig.~\ref{fig1}b). As such ensembles are typically subject to strong inhomogeneous broadening, an automatic separation into donor-like and acceptor like molecules will then take place. To quantify the emergent incoherent FRET migration behavior we provide a phenomenological model which allows analytical and numerical insight into the scaling of the VRS of molecular ensembles with density.\\
\subsection{FRET migration of excitation}
\indent For two near-field coupled adjacent molecules $j$ and $j'$, each with a single vibrational mode $b_j$ and $b_{j'}$ we perform a polaron transformation which leads to the following transformed operators $\tilde{\sigma}_j=\mathcal{Q}_j^\dagger\sigma_j$ and $\tilde{\sigma}_{j'}=\mathcal{Q}_{j'}^\dagger\sigma_{j'}$ (with displacement operators defined as $\mathcal{Q}_j=e^{\lambda_j (b_j^\dagger-b_j)}$). Under the assumption of low population of the excited electronic levels, the dipole-dipole interaction couples the quantum Langevin dynamics of the two molecules
\begin{subequations}
\begin{align}
\dot{\tilde{\sigma}}_j&=-\left(\gamma+i\delta_j\right)\tilde{\sigma}_j-i\Omega_{jj'}\tilde{\sigma}_{j'}\mathcal{Q}_{j'}\mathcal{Q}_j^\dagger+\sqrt{2\gamma}\tilde{\sigma}_j^{\text{in}},\\
\dot{\tilde{\sigma}}_{j'}&=-\left(\gamma+i\delta_{j'}\right) \tilde{\sigma}_{j'}-i\Omega_{jj'}\tilde{\sigma}_j\mathcal{Q}_j\mathcal{Q}_{j'}^\dagger+\sqrt{2\gamma}\tilde{\sigma}_{j'}^{\text{in}}.
\end{align}
\end{subequations}
\begin{figure*}[t]
\includegraphics[width=1.99\columnwidth]{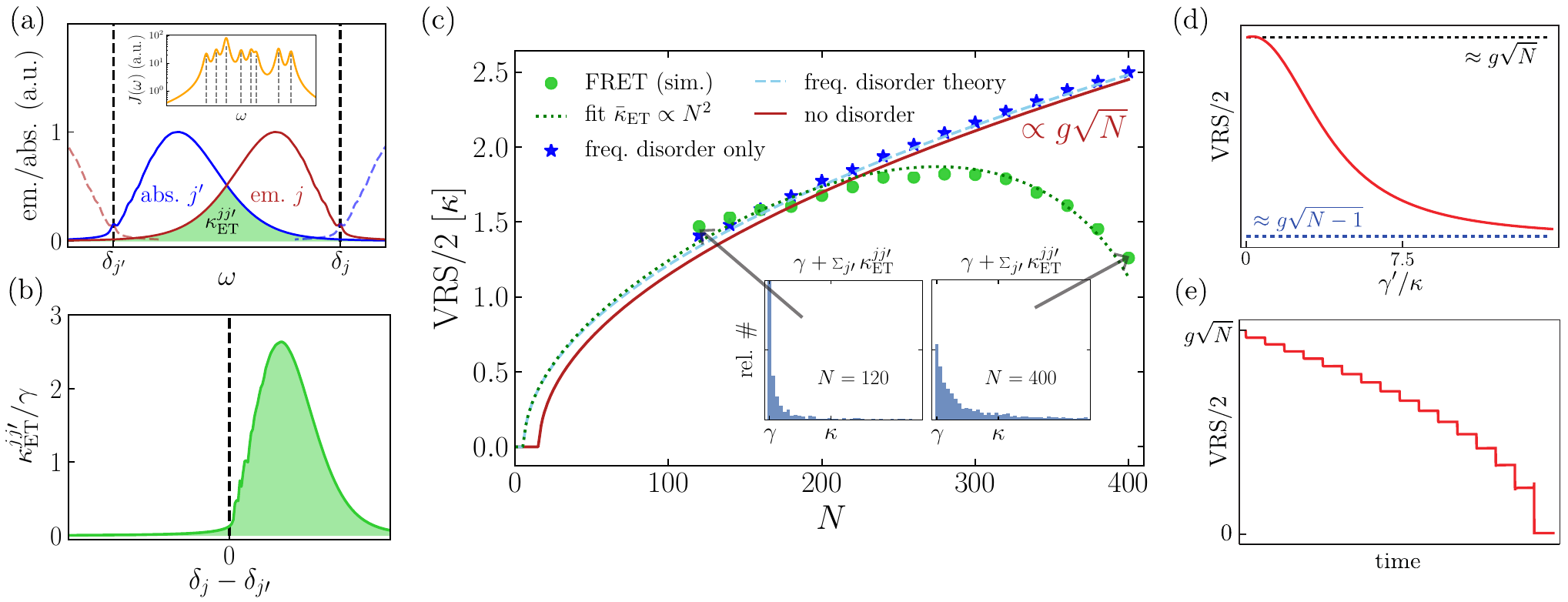}
\caption{a) Absorption and emission spectra for donor-acceptor pair with 8 vibrational modes (spectral density $J(\omega)$ shown in inset). The spectral overlap between $j$ and $j'$ gives rise to an incoherent rate $\kappa_{\text{ET}}^{jj'}$ from $j$ to $j'$. (b) Plot of $\kappa_{\text{ET}}^{jj'}$ as a function of frequency mismatch $\delta_j-\delta_{j'}$ for the same spectral density as in (a) and $\Omega_{jj'}=60\gamma$.  (c) Collective VRS for variable $\mathcal{N}$ in a sphere with radius $r=150\,\text{nm}$ for $n=1$. The parameters are $g=\kappa/8$, $\nu=0.5\kappa$, $w=\nu=0.5\kappa$, $\gamma=10^{-2}\kappa$, $\lambda=0.5$, $\Gamma=0.1\nu$. We averaged over 25 spatial realizations for each point. \C{The green dotted curve is a fit with $\bar{\kappa}_\text{ET}=3.16\times 10^{-5}\mathcal{N}^2\kappa$.} The blue stars show the results without FRET. The light blue dashed curve shows the (Markovian) theoretical prediction from Eq.~(\ref{theoryvrs}). The histograms show the normalized distributions of the decay rates $\gamma+\sum_j\kappa_{\text{ET}}^{jj\prime}$ for $\mathcal{N}=120$ and $\mathcal{N}=400$ particles, respectively. (d) Illustration of the gradual degradation of the VRS for $\mathcal{N}=16$ particles with increasing single strong donor behavior and $g=\kappa$. (e) Progressive degradation of the collective VRS for $\mathcal{N}=16$.}
\label{fig3}
\end{figure*}
This coupled dynamics can be solved in perturbation theory, assuming that $\Omega_{jj'}$ is small compared to the vibrational relaxation rates. The solution indicates an effective, largely unidirectional, energy transfer at rate $\kappa_\text{ET}^{jj'}$ from molecule $j$ to molecule $j'$. The rate is computed by assuming multiple paths of energy transfer between the two molecules involving all vibrational modes. We assume an initially electronically excited state with no vibrations present $\ket{e_j;0_1,0_2...0_{n}}$ of molecule $j$ and ground state without vibrations $\ket{g_{j'};0_1,0_2...0_{n}}$ for molecule $j'$. The emission of molecule $j$ leads it into state $\ket{g_j;m_1,...m_k,...m_{n}}$ and resonant interactions can occur with state $\ket{e_{j'};l_1,...l_{k'},...l_{n}}$ of molecule $j'$. Summing over all these processes leads to an analytical expression for the energy transfer rate (for detailed derivation see Appendix \ref{A6})
\begin{widetext}
\begin{equation}
\label{kappaET}
[\kappa_\text{ET}]^{jj'}=\sum_{m_1=0}^{\infty}\sum_{l_1=0}^{\infty}...\sum_{m_{n}=0}^{\infty}\sum_{l_{n}=0}^{\infty}     \left\{   \prod_{k=1}^{n}   e^{-2\lambda_k^2}    \frac{\lambda_k^{2(m_k+l_k)}}{m_k!{l}_k!}    \right\}       \frac{\sum_{k=1}^{n} (m_k+l_k) \Gamma_k \Omega_{jj'}^2 }{\left[\sum_{k=1}^{n} (m_k+l_k) \Gamma_k\right]^2+\left[\delta_j-\delta_{j'}-\sum_{k=1}^{n} (m_k+l_k) \nu_k\right]^2},
\end{equation}
\end{widetext}
which is the discrete version of the well established integral formulation~\cite{may2011charge} describing the overlap between the emission spectrum of molecule $j$ and absorption spectrum of molecule $j'$. This is illustrated in Fig.~\ref{fig3}a for a donor-acceptor pair with $8$ vibrational modes and spectral density $J(\omega)=\sum_k 2\lambda_k^2\nu_k^2\Gamma_k/[\Gamma_k^2+(\omega-\nu_k)^2]$ (stemming from the coupling of the molecular vibrations to some external phonon bath allowing for vibrational relaxation at rates $\Gamma_k$). The process is unidirectional as shown in Fig.~\ref{fig3}b as $\delta_j-\delta_{j'}$ dictates the direction of the energy flow.\\
\subsection{VRS scaling at high densities}

\indent The expression of the energy transfer rate in Eq.~\eqref{kappaET} greatly simplifies numerical simulations as it allows one to introduce an effective model of loss where for each pair of molecules $j$ and $j'$ a collapse operator $\sigma_j \sigma_{j'}^\dagger$ with corresponding rate $\kappa_\text{{ET}}^{jj'}$ is introduced. For molecule $j$, summing over all the paths of incoherent migration modifies the inherent radiative rate $\gamma$ to an increased one $\gamma+\textstyle \sum_{j'\neq j}^{\mathcal{N}}\kappa_\text{ET}^{jj'}$. Owing to the random spatial positioning within the ensemble, the FRET migration leads to an effective disorder in dissipation rates. This is directly incorporated in Eqs.~\eqref{eqsmotion} by amending the diagonal elements of the evolution matrix
\begin{equation}
M_{jj}=\gamma+i(\omega-\omega_\ell)+i\delta_j+\textstyle \sum_{j'\neq j}^{\mathcal{N}}\kappa_\text{ET}^{jj'}.
\end{equation}
Results are then possible for large systems (where a direct simulation of the evolution of the master equation is untractable) by a simple diagonalization of this matrix. The obtained scaling presented in Fig.~\ref{fig3}c shows that the FRET mechanism can lead to a strong deviation from the standard one ubiquitous in cavity QED with $\sqrt{\mathcal{N}}$.\\
\indent Beyond numerical estimates, a fully analytical approach based on the formalism introduced in the previous section  is possible allowing one to compute the effective polariton loss rate $\bar{\kappa}_\text{ET}+\tilde{\gamma}_{\text{dark}}$ stemming from the competition between static disorder and FRET migration. The average FRET dissipation rate
\begin{equation}
\bar{\kappa}_\text{ET}=\frac{1}{\mathcal{N}}\sum^{\mathcal{N}}_{j=1}\sum^{\mathcal{N}}_{j' \neq j}\kappa^{jj'}_{\text{ET}},
\end{equation}
is performed over the whole ensemble. In addition, the previously defined $\tilde{\gamma}_{\text{dark}}$ is derived from the matrix of Fourier transformed detunings and FRET rates (Appendix \ref{B2})
\begin{equation}
\tilde{\Delta}_{kk'} = \frac{1}{\mathcal{N}}\sum^{\mathcal{N}}_{j=1}\left(\delta_j - i\sum_{j'\neq j} \kappa_{\text{ET}}^{jj'}\right) e^{-i2\pi j(k-k')/\mathcal{N}},
\end{equation}
and it suffers modifications from the purely frequency disordered case. The new expression, seen as an extension of Eq.~\eqref{theoryvrs} (see Appendix~\ref{B2} for more details of the derivation) is
\begin{equation}
\label{App.D.Eq7b}
\text{VRS} \approx \Im\left\{2\sqrt{(\gamma + \bar{\kappa}_{\text{ET}} + \gamma_{\text{dark}}-\kappa)^2/4-g_\mathcal{N}^2} \right\}.
\end{equation}\\
 An averaging over disorder and position is possible assuming homogeneous media showing a scaling of $\bar{\kappa}_\text{ET}$ with $\mathcal{N}^2$. This allows for a fit of the result in Fig.~\ref{fig3}c showing that saturation stems from the strong increase of $\bar{\kappa}_\text{ET}$ with density squared.\\

\subsection{Reduction of VRS owing to strong donors}

\indent A very simple analysis in terms of bright and dark states can then shed insight into this scaling by assuming the effect of large decay onto the VRS. Assuming $\mathcal{N}-1$ molecules with identical decay rates $\gamma$ and a single lossier molecule with $\gamma'$ similar conclusions as in section \ref{Sec3}  referring to disorder are obtained. We can again cast the equations as
\begin{subequations}
\begin{align}
\dot{\beta}_1 &= -\gamma' \beta_1 - ig\alpha, \\
\dot{\mathcal{B}}_{\mathcal{N}-1} &= -\gamma\mathcal{B}_{\mathcal{N} -1} - ig\sqrt{\mathcal{N} -1}\alpha, \\
\dot{\alpha} &= -\kappa\alpha -ig\beta_1 -i\sqrt{\mathcal{N} -1}g\mathcal{B}_{\mathcal{N} -1}.
\end{align}
\end{subequations}

When $\gamma'=\gamma$ the system's bright state leads to polaritons at roughly $\pm g\sqrt{\mathcal{N}}$ while with increasing $\gamma'$ the polariton frequencies decrease to $\pm g\sqrt{\mathcal{N}-1}$ (as illustrated in Fig.~\ref{fig3}d). For the many strong donors case we illustrate in Fig.~\ref{fig3}e the gradual degradation of the VRS in time, from $g\sqrt{\mathcal{N}}$ to zero, when successively particles are turned from weak to strong donors. Following this interpretation, we added histograms in the inset of Fig.~\ref{fig3}c showing the distribution of dissipation rates within the ensemble. This provides a qualitative means to count out the number of lossy donors that fall out of the collective strong coupling condition.\\

\section{Conclusions and outlook}
\label{Sec5}
We proposed here an analytical approach which allows to quantify the effect of disorder on light-matter interactions in the strong coupling regime and which is extendable to molecular ensembles characterized by vibronic effects as well as near field effects owing to the electromagnetic vacuum. Employing a phenomenological model that incorporates all these aspects in an effective incoherent FRET migration of energy, scalings of the VRS with increasing density have been obtained, showing a strong divergence from the standard $\sqrt{\mathcal{N}}$ scaling in the absence of particle-particle interactions.\\
\indent The models used throughout the paper, albeit of limited validity, are standard and widely employed. For example, while the Holstein model for electron-vibron coupling is limited to molecules with large vibrational relaxation (such that anharmonicity is not reached) it provides a proper description to light-molecule~\cite{neuman2018origin,reitz2019langevin,wang2019turning} and molecule-molecule interactions~\cite{reitz2019langevin,reitz2020molecule}. The Tavis-Cummings model has also been universally used to predict and explain effects such as cavity mediated energy transfer~\cite{zhong2016non,zhong2017energy,feist2017long,reitz2019langevin}, energy and charge transport~\cite{orgiu2015conductivity,schachenmayer2015cavity,feist2015extraordinary,hagenmuller2017cavity,hagenmuller2018cavity,zeb2020incoherent}, cavity chemistry~\cite{galego2016suppressing,herrera2016cavity,herrera2020molecular,zhou2020polariton} etc. However, as some experiments are showing effects brought on by the onset of the ultrastrong coupling regime (such as for example the  work function of a material~\cite{hutchinson2013tuning}), recent theoretical works suggest that such regime is challenging and interesting~\cite{ridolfo2012photon,tuomas2016ultrastrong} and approaches are greatly interdisciplinary mixing aspects of quantum optics with quantum chemistry methods~\cite{haugland2020coupled}. In the future, we will extend our formalism based on linear quantum Langevin equations to include counter-rotating terms in the light-matter interaction Hamiltonian.

\acknowledgments
We acknowledge very useful discussions with Johannes Schachenmayer. We acknowledge financial support from the Max Planck Society and from the German Federal Ministry of Education and Research, co-funded by the European Commission (project RouTe), project number 13N14839 within the research program "Photonik Forschung Deutschland" (C.~G.). M.~R. acknowledges financial support from the International Max Planck Research School - Physics of Light (IMPRS-PL). This work was also funded by the Deutsche Forschungsgemeinschaft (DFG, German Research Foundation) -- Project-ID 429529648 -- TRR 306 QuCoLiMa (``Quantum Cooperativity of Light and Matter'').

\bibliography{Refs}

\onecolumngrid
\appendix

\section{The drift matrix}
\label{A2}
The drift matrix from section \ref{Sec3} of the main text reads (in the original basis)
\begin{eqnarray}
\label{Eq.CTrans1}
M = \left( \begin{array}{ccccc} (\omega - \omega_\ell + \delta_1 -i\gamma) & 0 & \cdots & 0 & g_1 \\ 0 & (\omega - \omega_\ell + \delta_2 -i\gamma) & \cdots & 0 & g_2 \\ \vdots & \vdots & \ddots & \vdots & \vdots \\ 0 & 0 & \cdots & (\omega - \omega_\ell + \delta_\mathcal{N}-i\gamma) & g_\mathcal{N} \\ g^{*}_1 & g^{*}_2 & \cdots & g^{*}_\mathcal{N} & (\omega_{c}-\omega_\ell -i\kappa) \end{array} \right).
\end{eqnarray}

\section{Elimination of the dark reservoir in the Markovian limit}
\label{A3}
\begin{figure}[t]
\includegraphics[width=0.8\columnwidth]{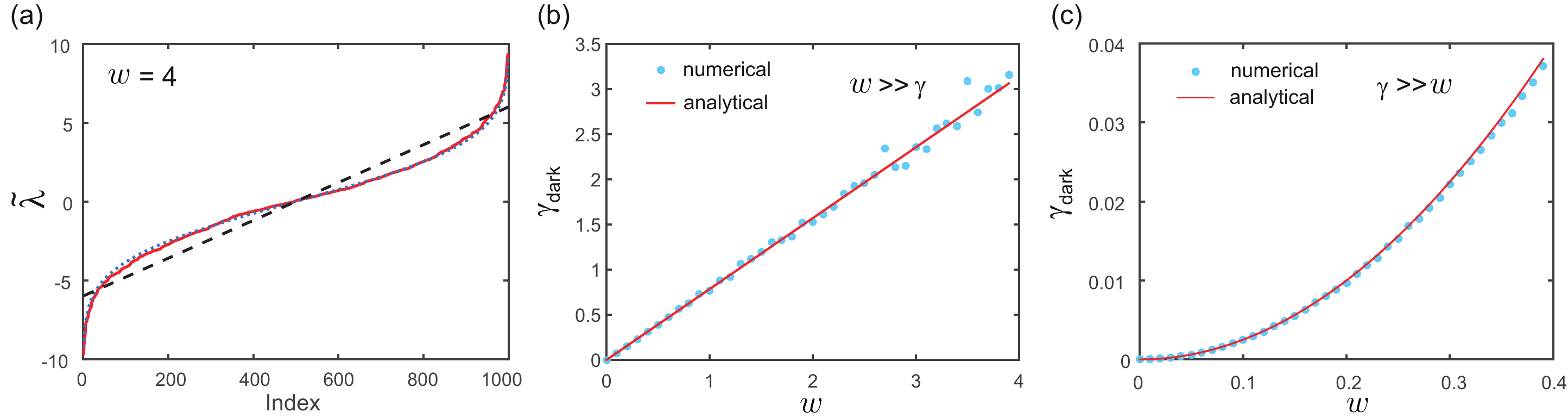}
\caption{(a) Sorted eigenvalue distribution for $\mathcal{N}=1000$ and $w=4$. The solid line shows the real part of the eigenvalues $\tilde{\lambda}_{1,\dots,\mathcal{N}-1}$ of the matrix $\mathcal{M}$ while the dashed line gives the linear eigenvalue distribution fit used in the derivation to determine $\gamma_{\text{dark}}$. The blue dotted line shows the quantile of the corresponding Gaussian distribution. (b) The blue dots give the numerical derivation of $\gamma_{\text{dark}}$ averaged over many disorder realizations (400) for $\mathcal{N}=100$ as a function of the width (standard deviation) $w$ of the disorder distribution. The solid red line shows the approximation with $\gamma_{\text{dark}} = (\pi/4)*w$. We have used $\gamma = 10^{-2}$. (c) In the case that $\gamma \gg w$ we find good agreement of the numerical data blue dots with the relation $\gamma_{\text{dark}} = w^2/\gamma$ given by the red line for $\gamma = 4$.}
\label{fig5}
\end{figure}

The expression for the frequency shift and decay rate induced by the dark state reservoir is:
\begin{equation}
\label{DrkRes.1}
\delta_\text{dark}+i \gamma_\text{dark}=\sum_{k,k'=1}^{\mathcal{N}-1}\Delta_{\mathcal{N}k}(\mathcal{M}^{-1})_{kk'}\Delta_{k'\mathcal{N}}=\frac{1}{\mathcal{N}^2}\sum_{j,j'=1}^{\mathcal{N}} \sum_{k,k'=1}^{\mathcal{N}-1} \delta_j\delta_{j'}(\mathcal{M}^{-1})_{kk'} e^{-i 2\pi (j k- j' k') /\mathcal{N}}.
\end{equation}
Some more insight can be obtained by evaluating the norm of the coupling vector for the dark states $\mathbf{m} = (\Delta_{1\mathcal{N}},\dots, \Delta_{\mathcal{N}-1 \mathcal{N}})^{\top}$. Here, we obtain
\begin{eqnarray}
\mathbf{m}^{\dagger}\mathbf{m} &=& \sum^{\mathcal{N}-1}_{k=1} |\Delta_{\mathcal{N}k}|^2 = \sum^{\mathcal{N}-1}_{k=1} \frac{1}{\mathcal{N}^2}\left(\sum^{\mathcal{N}}_{j,j'=1}\delta_{j} \delta_{j'} e^{i2\pi k(j-j')/\mathcal{N}} \right) = \frac{1}{\mathcal{N}^2}\sum^{\mathcal{N}}_{j,j'=1}\delta_{j} \delta_{j'} \left( \sum^{N}_{k=1} e^{i2\pi k(j-j')/\mathcal{N}}-1 \right) \\\nonumber
&=& \frac{1}{\mathcal{N}^2}\sum^{\mathcal{N}}_{j,j'=1}\delta_{j} \delta_{j'}(\mathcal{N}\delta_{jj'}-1)= \frac{1}{\mathcal{N}}\sum^{\mathcal{N}}_{j} \delta^{2}_{j} - \left(\frac{1}{\mathcal{N}}\sum^{\mathcal{N}}_{j} \delta_j\right)^2 = \text{Var}(\delta),
\end{eqnarray}
which is the variance of the frequency distribution. In the case of a Gaussian distribution $p(\delta) = (1/\sqrt{2\pi w^2})e^{-\delta^2/(2w^2)}$ for the disorder we obtain $\text{Var}(\delta) = w^2$. We can use this result to rewrite $\mathbf{m} = w \hat{\mathbf{m}}$, where $\hat{\mathbf{m}}$ is the normalized vector of $\mathbf{m}$. This allows us to further evaluate the expression in Eq.~\ref{DrkRes.1} to
\begin{eqnarray}
\delta_\text{dark}+i \gamma_\text{dark} = \mathbf{m}^{\dagger}\mathcal{M}^{-1}\mathbf{m}
= w^2 \hat{\mathbf{m}}^{\dagger}\mathcal{M}^{-1}\hat{\mathbf{m}}
= w^2 \hat{\mathbf{c}}^{\dagger} \tilde{D}^{-1} \hat{\mathbf{c}},
\end{eqnarray}
where $\mathcal{M} = T\tilde{D}T^{\dagger}$ with the diagonal matrix $\tilde{D}$ and $\hat{\mathbf{c}} = T^{\dagger}\hat{\mathbf{m}}$. Since $\tilde{D}^{-1}$ is diagonal, we can find the expression
\begin{eqnarray}
\delta_\text{dark}+i \gamma_\text{dark} &=& w^2\sum^{\mathcal{N}-1}_{j=1} |\hat{c}_j|^2\tilde{D}^{-1}_{jj} \approx  w^2 \sum^{\mathcal{N}-1}_{j=1}|\hat{c}_j|^2  \tilde{\lambda}^{-1}_j,
\end{eqnarray}
which is the weighted average over the eigenvalues $\tilde{\lambda}^{-1}_j$ of the matrix $\mathcal{M}^{-1}$ times the variance of the distribution. Since the eigenvalues of $\mathcal{M}$ follow the form $\lambda_{1,\dots,\mathcal{N}-1} = \bar{\delta}-i\gamma - \tilde{\lambda}_{1,\dots,N-1}$ where $\tilde{\lambda}_{1,\dots,\mathcal{N}-1} \in \mathbb{R}$ we obtain
\begin{equation}
\delta_\text{dark}+i \gamma_\text{dark} = w^2\sum^{\mathcal{N}-1}_{j=1}|\hat{c}_j|^2 \frac{1}{\bar{\delta}-i\gamma - \tilde{\lambda}_{j}}.
\end{equation}
For large $\mathcal{N}$ where $\bar{\delta} \rightarrow 0$ we finally obtain the expression
\begin{eqnarray}
\delta_\text{dark}+i \gamma_\text{dark} &=& w^2\sum^{\mathcal{N}-1}_{j=1}|\hat{c}_j|^2 \frac{-1}{\tilde{\lambda}_{j} + i\gamma}
= \left(w^2\sum^{\mathcal{N}-1}_{j=1}|\hat{c}_j|^2\frac{-\tilde{\lambda}_{j}}{\tilde{\lambda}^2_{j} + \gamma^2}\right) + i\left(w^2\sum^{\mathcal{N}-1}_{j=1}|\hat{c}_j|^2 \frac{\gamma}{\tilde{\lambda}^2_{j} + \gamma^2}\right).
\end{eqnarray}
In the case that $\gamma \gg \tilde{\lambda}_{1,\dots,\mathcal{N}-1}$ which is given if $\gamma \gg w$, we can obtain a solution for $\gamma_{\text{dark}}$ which is obtained from
\begin{eqnarray}
\delta_\text{dark}+i \gamma_\text{dark} \approx \left(w^2\sum^{\mathcal{N}-1}_{j=1}|\hat{c}_j|^2\frac{-\tilde{\lambda}_{j}}{\gamma^2}\right) + i\left(w^2\sum^{\mathcal{N}-1}_{j=1}|\hat{c}_j|^2 \frac{1}{\gamma}\right)
= \left(-\frac{w^2}{\gamma^2}\sum^{\mathcal{N}-1}_{j=1}|\hat{c}_j|^2\tilde{\lambda}_{j}\right) + i\left(\frac{w^2}{\gamma}\right),
\end{eqnarray}
since $\sum^{\mathcal{N}-1}_{j=1}|\hat{c}_j|^2 = 1$. Since the eigenvalues $\tilde{\lambda}_{1,\dots,\mathcal{N}-1}$ are equally distributed around zero as shown in Fig.~\ref{fig4}a and the weights can be roughly approximated by $|\hat{c}_j|^2 \approx 1/(\mathcal{N}-1)$ (due to the fact that $\braket{|\hat{c}_j|^2} = 1/(\mathcal{N}-1)$), the frequency shift becomes $\delta_{\text{dark}} \rightarrow 0$ for large $\mathcal{N}$ and $\gamma_{\text{dark}} = w^2/\gamma$.\\
A rough approximation can be performed in the case $\tilde{\lambda}_{j} \gg \gamma$ for most $j \in \{1,\dots,\mathcal{N}-1\}$. Here we set again $|\hat{c}_j|^2 \approx 1/(\mathcal{N}-1)$ and we obtain
\begin{eqnarray}
\delta_\text{dark}+i \gamma_\text{dark} &\approx& \left(\frac{w^2}{\mathcal{N}-1}\sum^{\mathcal{N}-1}_{j=1}\frac{-\tilde{\lambda}_{j}}{\tilde{\lambda}^2_{j} + \gamma^2}\right) + i\left(\frac{w^2}{\mathcal{N}-1}\sum^{\mathcal{N}-1}_{j=1}\frac{\gamma}{\tilde{\lambda}^2_{j} + \gamma^2}\right) \\\nonumber
&\approx & -\frac{w}{2q}\int^{qw}_{-qw}d\tilde{\lambda} \frac{\tilde{\lambda}}{\tilde{\lambda}^2 + \gamma^2}  + i\frac{w\gamma}{2q}\int^{qw}_{-qw} d\tilde{\lambda} \frac{1}{\tilde{\lambda}^2 + \gamma^2}
\approx  i\frac{w\gamma}{2q}\left[\frac{2}{\gamma}\arctan\left(\frac{qw}{\gamma} \right) \right]
\approx  i\frac{\pi}{2q}w,
\end{eqnarray}
where we have assumed that the eigenvalues $\tilde{\lambda}_j$ are linearly distributed from $-qw$ to $qw$ where $q$ is an adjustment or fitting parameter. This is a rather rough approximation of the real eigenvalue distribution depicted in Fig.~\ref{fig5}a. In reality the sorted eigenvalue distribution follows $\tilde{\lambda}_j = \sqrt{2\pi}w\erf^{-1}(2j/\mathcal{N}-1)$ for $j\in \{1,\dots, \mathcal{N}-1\}$ when $\mathcal{N} \rightarrow \infty$, which is the quantile function of the Gaussian distribution for the energy disorder. The best approximation is given for $q=2$ resulting in $\gamma_{\text{dark}} = (\pi/4)w$ as shown in Fig.~\ref{fig5}b while $\delta_{\text{dark}} = 0$.
\section{Quantum Langevin equations approach to non-Markovian loss into the dark reservoir}
\label{A4}
Starting with the full equations of motion for the bright and dark modes given by
\begin{subequations}
\begin{align}
\label{DrkMark1}
\dot{\hat{\mathbf{v}}}&= -i\mathcal{M}\hat{\mathbf{v}} - i\mathbf{m} \hat{B} + \sqrt{2\gamma}\hat{\mathbf{v}}_{\text{in}},\\
\label{DrkMark2}
\dot{\hat{B}} &= -i( \bar{\delta} -i\gamma)\hat{B}-i\mathbf{m}^{\dagger}\hat{\mathbf{v}}  + \sqrt{2\gamma}\hat{B}_{\text{in}},
\end{align}
\end{subequations}
where we have defined $\hat{\mathbf{v}} = (\hat{D}_{1}, \dots, \hat{D}_{\mathcal{N}-1})^{\top}$ we obtain by injecting the steady state solution for the dark modes
\begin{eqnarray}
\hat{\mathbf{v}}(t) &=& \int^{t}_{-\infty}ds e^{-i\mathcal{M}(t-s)}\left(-i\mathbf{m}\hat{B}(s) + \sqrt{2\gamma}\hat{\mathbf{v}}_{\text{in}}(s) \right),
\end{eqnarray}
into Eq.~\eqref{DrkMark2} the reduced equation of motion for the bright mode
\begin{eqnarray}
\nonumber
\dot{\hat{B}} &=& -i( \bar{\delta} -i\gamma)\hat{B} - \int^{\infty}_{-\infty}ds \Theta(t-s) \mathbf{m}^{\dagger}e^{-i\mathcal{M}(t-s)}\mathbf{m}\hat{B}(s) -i\sqrt{2\gamma}\int^{\infty}_{-\infty}ds \Theta(t-s) \mathbf{m}^{\dagger}e^{-i\mathcal{M}(t-s)}\hat{\mathbf{v}}_{\text{in}}(s) + \sqrt{2\gamma}\hat{B}_{\text{in}} \\
&=& -i( \bar{\delta} -i\gamma)\hat{B} - \int^{\infty}_{-\infty}ds \Theta(t-s) \sum^{\mathcal{N}-1}_{j=1}|c_j|^2 e^{-i(\bar{\delta}-i\gamma - \tilde{\lambda}_j)(t-s)}\hat{B}(s) + \xi_{\text{dark}} + \sqrt{2\gamma}\hat{B}_{\text{in}},
\end{eqnarray}
where we have defined the noise term emerging from the dark reservoir by $\xi_{\text{dark}}(t) =  -i\sqrt{2\gamma}\int^{\infty}_{-\infty}ds \Theta(t-s) \mathbf{m}^{\dagger}e^{-i\mathcal{M}(t-s)}\hat{\mathbf{v}}_{\text{in}}(s)$. Following the same steps introduced in the previous subsection where we approximate $|c_j|^2 \approx w^2/(\mathcal{N}-1) $ and assuming $\tilde{\lambda}_j$ being linearly distributed between $-2w$ and $2w$ we obtain
\begin{eqnarray}
\sum^{\mathcal{N}-1}_{j=1}|c_j|^2 e^{i\tilde{\lambda}_j(t-s)} \approx  \frac{w}{4}\int^{2w}_{-2w}d\tilde{\lambda} e^{i\tilde{\lambda}(t-s)}
= w^2\sinc(2w(t-s)),
\end{eqnarray}
for the convolution kernel which results in
\begin{equation}
\label{DrkMark3}
\dot{\hat{B}} =  -i( \bar{\delta} -i\gamma)\hat{B} -w^2 \int^{\infty}_{-\infty}ds \Theta(t-s) e^{-i(\bar{\delta}-i\gamma)(t-s)} \sinc(2w(t-s)) \hat{B}(s) + \xi_{\text{dark}} + \sqrt{2\gamma}\hat{B}_{\text{in}}.
\end{equation}
Before taking the limit for large width $w \gg \gamma$, we evaluate the noise correlation term for the dark states
\begin{eqnarray}
\langle \xi_{\text{dark}}(t) \xi^{\dagger}_{\text{dark}}(t') \rangle &=& 2\gamma\int^{\infty}_{-\infty}ds \int^{\infty}_{-\infty}ds'\Theta(t-s)\Theta(t'-s') \mathbf{m}^{\dagger} e^{-i\mathcal{M}(t-s)}\langle \hat{\mathbf{v}}_{\text{in}}(s) \hat{\mathbf{v}}_{\text{in}}(s') \rangle e^{i\mathcal{M}^{\dagger}(t'-s')}\mathbf{m} \\\nonumber
&=& 2\gamma\int^{\infty}_{-\infty}ds \int^{\infty}_{-\infty}ds'\Theta(t-s)\Theta(t'-s') \mathbf{m}^{\dagger} e^{-i\mathcal{M}(t-s)}\delta(s-s')\mathds{1} e^{i\mathcal{M}^{\dagger}(t'-s')}\mathbf{m} \\\nonumber
&=& 2\gamma\int^{\infty}_{-\infty}ds \Theta(t-s)\Theta(t'-s) \mathbf{m}^{\dagger} e^{-i\mathcal{M}(t-s)}e^{i\mathcal{M}^{\dagger}(t'-s)}\mathbf{m} \\\nonumber
&=& 2\gamma^{-i\bar{\delta}(t-t')}e^{-\gamma(t+t')}\mathbf{m}^{\dagger} e^{-i\hat{\Delta}(t-t')}\mathbf{m}\int^{\infty}_{-\infty}ds \Theta(t-s)\Theta(t'-s) e^{2\gamma s} \\\nonumber
&=& \left\{\begin{array}{cc} \mathbf{m}^{\dagger}e^{-i\mathcal{M}(t-t')}\mathbf{m} = \sum^{\mathcal{N}-1}_{j=1}|c_j|^2 e^{-i(\bar{\delta}-i\gamma - \tilde{\lambda}_j)(t-t')}, & t \geq t' \\  \mathbf{m}^{\dagger}e^{i\mathcal{M}^{\dagger}(t'-t)}\mathbf{m} = \sum^{\mathcal{N}-1}_{j=1}|c_j|^2 e^{i(\bar{\delta}+i\gamma - \tilde{\lambda}_j)(t'-t)}, & t < t'. \end{array} \right.
\end{eqnarray}
Here, we have used the definition $\hat{\Delta} = \mathcal{M} + i\gamma\mathds{1}$.
For our approximation, this equates to
\begin{equation}
\langle \xi_{\text{dark}}(t) \xi^{\dagger}_{\text{dark}}(t') \rangle = \left\{ \begin{array}{cc} w^2 e^{-i(\bar{\delta}-i\gamma)(t-t')}\sinc(2w(t-t')), & t \geq t' \\ w^2 e^{i(\bar{\delta}+i\gamma)(t'-t)}\sinc(2w(t'-t)), & t < t'  \end{array} \right.
\end{equation}
For very large width $w \gg \gamma$ we can make the approximation $\sin(2w(t-t'))/(\pi(t-t')) \approx \delta(t-t')$ which results in
\begin{eqnarray}
\langle \xi_{\text{dark}}(t) \xi^{\dagger}_{\text{dark}}(t') \rangle  \approx  \frac{\pi}{2}w\delta(t-t')
= 2\gamma_{\text{dark}}\delta(t-t'),
\end{eqnarray}
which describes delta correlated noise (Markovian noise) in this limit. With the correlation relation $2\gamma\langle \hat{B}_{\text{in}}(t)\hat{B}_{\text{in}}(t')\rangle = 2\gamma \delta(t-t')$ for the noise of the bright mode and simplifying the term
\begin{eqnarray}
w^2 \int^{\infty}_{-\infty}ds \Theta(t-s) e^{-i(\bar{\delta}-i\gamma)(t-s)} \sinc(2w(t-s)) \hat{B}(s) &=& \frac{\pi}{2}w \int^{\infty}_{-\infty}ds \Theta(t-s) e^{-i(\bar{\delta}-i\gamma)(t-s)} \frac{\sin(2w(t-s))}{\pi(t-s)} \hat{B}(s) \\\nonumber
&\approx&  \frac{\pi}{2}w \int^{\infty}_{-\infty}ds \Theta(t-s) e^{-i(\bar{\delta}-i\gamma)(t-s)} \delta(t-s) \hat{B}(s) \\\nonumber
&=&\frac{\pi}{4}w\hat{B}(t)
= \gamma_{\text{dark}} \hat{B}(t),
\end{eqnarray}
in Eq.~\eqref{DrkMark3}, we obtain for the equation of motion in the Markovian limit the expression
\begin{equation}
\label{DrkMark4}
\dot{\hat{B}} = -i\left(\bar{\delta} -i\left(\gamma + \gamma_{\text{dark}}\right)\right)\hat{B} + \sqrt{2\left(\gamma + \gamma_{\text{dark}}\right)}\hat{B}_{\text{in}}.
\end{equation}
The Markovian limit can be obtained much more directly and irrespective of the given shape of the disorder distribution in the case where $\gamma \gg w$. By using the relation $(\gamma/2)\exp(-\gamma|t-s|) \approx \delta(t-s)$ for large $\gamma$ we can
rewrite
\begin{eqnarray}
 \int^{\infty}_{-\infty}ds \Theta(t-s) \sum^{\mathcal{N}-1}_{j=1}|c_j|^2 e^{-i(\bar{\delta}-i\gamma - \tilde{\lambda}_j)(t-s)}\hat{B}(s) & \approx & \frac{1}{\gamma}\sum^{\mathcal{N}-1}_{j=1} |c_j|^2\hat{B}(t)  =   \frac{\text{Var}(\delta)}{\gamma}\hat{B}(t),
\end{eqnarray}
while simultaneously we obtain for the noise correlation term
\begin{eqnarray}
\langle \xi_{\text{dark}}(t) \xi^{\dagger}_{\text{dark}}(t') \rangle & = & \left\{\begin{array}{cc} e^{-i(\bar{\delta}-i\gamma)(t-t')}\sum^{\mathcal{N}-1}_{j=1}|c_j|^2 e^{i\tilde{\lambda}_j (t-t')}, & t \geq t' \\  e^{i(\bar{\delta}+i\gamma)(t'-t)}\sum^{\mathcal{N}-1}_{j=1}|c_j|^2 e^{-i\tilde{\lambda}_j (t'-t)}, & t < t' \end{array} \right. \\\nonumber
& \approx & \frac{2}{\gamma}\left(\sum^{\mathcal{N}-1}_{j=1}|c_j|^2\right)\delta(t-t')
 =  2\left(\frac{\text{Var}(\delta)}{\gamma}\right)\delta(t-t').
\end{eqnarray}
This results in $\gamma_{\text{dark}} = \text{Var}(\delta)/\gamma$ for Eq.~\eqref{DrkMark4}. In this regime in particular we have identified
\begin{equation}
\hat{B}_{\text{in}}(t) = -i\frac{\mathbf{m}^{\dagger}\mathbf{v}_{\text{in}}(t)}{\sqrt{\mathbf{m}^{\dagger}\mathbf{m}}}.
\end{equation}

\section{Orientational disorder}
\label{A5}
In the case that we have molecules with randomly oriented dipole moments the cavity coupling strength $g_j$ varies from molecule to molecule. The effect of this disorder manifest itself in the definition of the bright and dark modes where the bright mode is now given by $\hat{B} = \left(1/\sqrt{\sum_{j=1}^{\mathcal{N}}|g_j|^2}\right)\sum_{j=1}^{\mathcal{N}} g^{*}_j \sigma_j$ which results in the equation of motion for the cavity
\begin{equation}
\dot{\alpha} = -i(\delta - i\kappa)\alpha - i\tilde{g}_{\mathcal{N}}\mathcal{B} + \eta,
\end{equation}
where $\tilde{g}_\mathcal{N}= \sqrt{\sum_{j=1}^{\mathcal{N}}|g_j|^2}$ and $\delta = \omega_c - \omega_l$. Taking the limit for large numbers of molecules $\mathcal{N}$ where we describe the random orientation with respect to the electric field of the cavity mode $a$ by $g_j = g\cos(\theta_j)$ where $\theta_j \in [0, \pi]$ we obtain
\begin{eqnarray}
\tilde{g}^2_\mathcal{N} &\approx & \frac{\mathcal{N}g^2}{\pi}\int^{\pi}_{0}d\theta \cos^2(\theta) = \frac{\mathcal{N}g^2}{2}.
\end{eqnarray}
The dark modes $\hat{D}_k = \left(1/\sqrt{\sum_{j=1}^{\mathcal{N}}|g_j|^2}\right)\sum_{j=1}^{N} d^{*}_{kj} \sigma_j$ can be obtained by employing the Gram-Schmidt orthogonalization procedure starting with the vector $\left(1/\sqrt{\sum_{j=1}^{\mathcal{N}}|g_j|^2}\right)(g^{*}_1,\dots , g^{*}_\mathcal{N})$ containing the coefficients of the bright mode. This results in
\begin{subequations}
\begin{align}
\dot{\mathcal{D}}_j &= -i(\delta_l + \Delta^{(g)}_{jj} - i\gamma)\mathcal{D}_j  -i\sum^{\mathcal{N}-1}_{j \neq j'}\Delta^{(g)}_{jj'}\mathcal{D}_{j'} -i\Delta^{(g)}_{j\mathcal{N}}\mathcal{B}, \\
\dot{\mathcal{B}} &= -i(\delta_l + \delta_{\mathcal{B}} - i\gamma)\mathcal{B}  -i\sum^{\mathcal{N}-1}_{j'=1}\Delta^{(g)}_{\mathcal{N}j'}\mathcal{D}_{j'} -i\tilde{g}_{N}\alpha, \\
\dot{\alpha} &= -i(\delta - i\kappa)\alpha - i\tilde{g}_{\mathcal{N}}\mathcal{B} + \eta,
\end{align}
\end{subequations}
where $\Delta^{(g)}_{jj'} = \left(\sum_{j=1}^{\mathcal{N}}|g_j|^2\right)^{-1}\sum^{\mathcal{N}}_{k=1}d^{*}_{jk}\delta_k d_{kj'}$, $\delta_l = \omega - \omega_l$ and $\delta_{\mathcal{B}} = \Delta^{(g)}_{\mathcal{N}\mathcal{N}} = \left(\sum_{j=1}^{\mathcal{N}}|g_j|^2\right)^{-1}\sum^{\mathcal{N}}_{j=1}\delta_j|g_j|^2$.
From the equations of motion we obtain the matrix for the dark states
\begin{equation}
\mathcal{M}^{(g)} =\left(
    \begin{array}{cccc}
      (\delta_l + \delta_{\mathcal{D}_1}-i\gamma) & \Delta^{(g)}_{12} & \ldots & \Delta^{(g)}_{1(\mathcal{N}-1)} \\
      \Delta^{(g)}_{21} & (\delta_l + \delta_{\mathcal{D}_2}-i\gamma) & \ldots & \Delta^{(g)}_{2(\mathcal{N}-1)}  \\
      \vdots & \vdots & \ddots & \vdots \\
      \Delta^{(g)}_{(\mathcal{N}-1)1} & \Delta^{(g)}_{(\mathcal{N}-1)2} & \ldots & (\delta_l + \delta_{\mathcal{D}_{\mathcal{N}-1}}-i\gamma) \\
    \end{array}
  \right),
\end{equation}
where $\delta_{\mathcal{D}_k} = \Delta^{(g)}_{kk}$. Assuming steady state for the dark manifold we obtain the reduced equations of motion
\begin{subequations}
\begin{align}
\dot{\mathcal{B}} &= -i\left(\delta_l + \delta_{\mathcal{B}} -\sum^{\mathcal{N}-1}_{kk'=1}\Delta^{(g)}_{\mathcal{N}k}\left[(\mathcal{M}^{(g)})^{-1}\right]_{kk'}\Delta^{(g)}_{k'\mathcal{N}}  - i\gamma\right)\mathcal{B} -i\tilde{g}_{\mathcal{N}}\alpha, \\
\dot{\alpha} &= -i(\delta - i\kappa)\alpha - i\tilde{g}_{\mathcal{N}}\mathcal{B} + \eta.
\end{align}
\end{subequations}

\section{FRET migration process rates}
\label{A6}
Here we want to derive a first-order FRET rate between two near-field coupled molecules $j$ and $j'$, each with a single vibrational mode $b_j$ and $b_{j'}$ \cite{reitz2019langevin}. To this end, we go into a polaron frame and start with the  equations of motion for the dressed dipole operators $\tilde{\sigma}_j=\mathcal{Q}_j^\dagger\sigma_j$ and $\tilde{\sigma}_{j'}=\mathcal{Q}_{j'}^\dagger\sigma_{j'}$ with the displacement operator $\mathcal{Q}_j=e^{\lambda_j (b_j^\dagger-b_j)}$:
\begin{subequations}
\label{eom}
\begin{align}
\dot{\tilde{\sigma}}_j&=-\left(\gamma+i\delta_j\right)\tilde{\sigma}_j-i\Omega_{jj'}\tilde{\sigma}_{j'}\mathcal{Q}_{j'}\mathcal{Q}_j^\dagger+\sqrt{2\gamma}\tilde{\sigma}_j^{\text{in}},\\
\dot{\tilde{\sigma}}_{j'}&=-\left(\gamma+i\delta_{j'}\right) \tilde{\sigma}_{j'}-i\Omega_{jj'}\tilde{\sigma}_j\mathcal{Q}_j\mathcal{Q}_{j'}^\dagger+\sqrt{2\gamma}\tilde{\sigma}_{j'}^{\text{in}}.
\end{align}
\end{subequations}
For simplicity, for the numerical simulations, we will assume parallel orientation of all dipoles. The dipole-dipole interaction then expresses as $\Omega_{jj'}=\frac{3}{2}\gamma/(k|\mathbf{r}_j-\mathbf{r}_{j'}|)^3$ with wavenumber $k=2\pi/\lambda_0$ ($\lambda_0$ is the wavelength of the electronic transition) \cite{lehmberg1970radiation}. We will consider initial excitation of molecule $j$ and assume molecule $j'$ to be in the electronic ground state initially. The equation of motion for the acceptor's population reads:
\begin{align}
\dot{P}_{j'}=-2\gamma P_{j'}+2\Omega_{jj'}\Im\braket{\sigma_{j'}^\dagger\sigma_j}+\sqrt{2\gamma}\braket{\sigma_{j'}^\dagger\sigma_{j'}^{\text{in}}+\sigma_{j'}^{\dagger,\text{in}}\sigma_{j'}}.
\end{align}
We therefore have to evaluate the term $2\Omega_{jj'}\Im\braket{\sigma_{j'}^\dagger\sigma_j}$ which signals the energy transfer. Formal integration of the equation of motion for the acceptor gives
\begin{align}
\sigma_{j'}^\dagger (t)=\sigma_{j'}(0)e^{-(\gamma-i\delta_{j'}) t}\mathcal{Q}_{j'}(0)\mathcal{Q}_{j'}^\dagger (t)+\int_{0}^t dt' e^{-(\gamma-i\delta_{j'})(t-t')}\left[i\Omega_{jj'}\sigma_j^\dagger(t')+\sqrt{2\gamma}\sigma_{j'}^{\dagger,\text{in}}(t')\right]\mathcal{Q}_{j'} (t')\mathcal{Q}_{j'}^\dagger (t).
\end{align}
The correlation $\braket{\sigma_{j'}^\dagger(t)\sigma_j(t)}$ can then be expressed  as (assuming free evolution of $j$)
\begin{align}
\braket{\sigma_{j'}^\dagger\sigma_j}=i\Omega_{jj'}\int_0^t dt' e^{-(\gamma+i(\delta_j-\delta_{j'}))(t-t')}e^{-\gamma t'}e^{-\gamma t} P_j (0)\braket{\mathcal{Q}_j (0)\mathcal{Q}_j^\dagger (t')\mathcal{Q}_j(t)\mathcal{Q}_j^\dagger(0)}\braket{\mathcal{Q}_{j'}(t')\mathcal{Q}_{j'}^\dagger (t)}
=:i\Omega_{jj'} P_j (0)\cdot \mathcal{I}(t).
\end{align}
where we defined $P_j(0)=\braket{\sigma_j^\dagger\sigma_j (0)}$. We therefore have to evaluate the two correlation functions $\braket{\mathcal{Q}_{j'}(t')\mathcal{Q}_{j'}^\dagger (t)}$ and $\braket{\mathcal{Q}_j (0)\mathcal{Q}_j^\dagger (t')\mathcal{Q}_j(t)\mathcal{Q}_j^\dagger(0)}$ (we will assume identical Huang-Rhys factors for both molecules $\lambda:=\lambda_j=\lambda_{j'}$):
\begin{align}
\braket{\mathcal{Q}_{j'}(t')\mathcal{Q}_{j'}^\dagger (t)}&=e^{-\lambda^2}e^{\lambda^2e^{-(\Gamma-i\nu)(t-t')}},\\
\braket{\mathcal{Q}_j (0)\mathcal{Q}_j^\dagger (t')\mathcal{Q}_j(t)\mathcal{Q}_j^\dagger(0)}&=e^{-\lambda^2}e^{\lambda^2 e^{-(\Gamma-i\nu)(t-t')}}e^{-\lambda^2 e^{-(\Gamma+i\nu)t'}}e^{\lambda^2 e^{-(\Gamma-i\nu)t'}},
\end{align}
where to obtain the expression for the latter one, we commuted $\mathcal{Q}_j^\dagger(0)$ with $\mathcal{Q}_j(t)$ and $\mathcal{Q}_j^\dagger (t')$, respectively and assumed large times $t\gg 1/\Gamma$. We can now evaluate the integral $\mathcal{I}(t)$:
\begin{align}
\mathcal{I}(t)=e^{2\lambda^2}\sum_{n_1,n_2,n_3,n_4}s_{n_1}^\lambda s_{n_2}^\lambda s_{n_3}^\lambda s_{n_4}^\lambda (-1)^{n_2}\frac{e^{-[2\gamma+(n_2+n_3)\Gamma+i(n_2-n_3)\nu]t}-e^{[-2\gamma-(n_1+n_4)\Gamma-i(\Delta-n_1\nu-n_4\nu)]t}}{(n_1-n_2+n_3+n_4)\Gamma+i(\delta_j-\delta_{j'}-n_1-n_2+n_3-n_4)\nu}.
\end{align}
Due to the fast decay of terms containing $\Gamma$ (we assume $\Gamma\gg\gamma$), we can approximate the energy transfer rate as ($n_2=n_3=0$)
\begin{align}
2\Omega\braket{\sigma_{j'}^\dagger\sigma_j}&=\sum_{n_1,n_4}\frac{2\Omega_{jj'}^2 s_{n_1}^\lambda s_{n_4}^\lambda(n_1+n_4)\Gamma}{(n_1+n_4)^2\Gamma^2+[\delta_j-\delta_{j'}-(n_1+n_4)\nu]^2}P_j(0)e^{-2\gamma t}\\\nonumber
&\approx\sum_{n_1,n_4}\frac{2\Omega_{jj'}^2 s_{n_1}^\lambda s_{n_4}^\lambda(n_1+n_4)\Gamma}{(n_1+n_4)^2\Gamma^2+[\delta_j-\delta_{j'}-(n_1+n_4)\nu]^2}P_j(t)\\\nonumber
&=\sum_{n_j,n_{j'}}\frac{2\Omega_{jj'}^2 s_{n_j}^\lambda s_{n_{j'}}^\lambda(n_j+n_{j'})\Gamma}{(n_j+n_{j'})^2\Gamma^2+[\delta_j-\delta_{j'}-(n_{j}+n_{j'})\nu]^2}P_j(t)=:2[\kappa_{\text{ET}}]^{jj'}P_j(t),
\end{align}
where we introduced the Poissonian coefficients $s_{n}^\lambda=e^{-\lambda^2}\lambda^{2n}/n!$.\\

\indent In the case of many vibrational modes $n$ for donor and acceptor, we can generalize the result by writing general displacements $\mathcal{Q}_{j'}=\prod_{k=1}^{n} \mathcal{Q}_{j'}^k$ and $\mathcal{Q}_j=\prod_{k=1}^{n} \mathcal{Q}_j^k$ for all vibrational modes. The equations of motion can then be expressed in the same form as in Eqs.~(\ref{eom})
\begin{subequations}
\begin{align}
\dot{\tilde{\sigma}}_j&=-\left(\gamma+i\delta_j\right)\tilde{\sigma}_j-i\Omega\tilde{\sigma}_{j'}\mathcal{Q}_{j'}\mathcal{Q}_j^\dagger+\sqrt{2\gamma}\tilde{\sigma}_j^{\text{in}},\\
\dot{\tilde{\sigma}}_{j'}&=-\left(\gamma+i\delta_{j'}\right) \tilde{\sigma}_{j'}-i\Omega\tilde{\sigma}_j\mathcal{Q}_j\mathcal{Q}_{j'}^\dagger+\sqrt{2\gamma}\tilde{\sigma}_{j'}^{\text{in}}.
\end{align}
\end{subequations}
We further more assume that different vibrational modes are independent of each other, i.e., we assume factorizability of all correlation functions, e.g.~
\begin{align}
\braket{\mathcal{Q}_{j'}(t')\mathcal{Q}_{j'}^\dagger (t)}=\prod_k\braket{\mathcal{Q}_{j'}^k(t')\mathcal{Q}_{j'}^{k,\dagger} (t)}=\prod_k e^{-\lambda_k^2}e^{\lambda_k^2 e^{-(\Gamma_k-i\nu_k)(t-t')}}.
\end{align}
Assuming that the two molecules have the same vibrational properties $\braket{\mathcal{Q}_{j'}^k(t')\mathcal{Q}_{j'}^{k,\dagger} (t)}=\braket{\mathcal{Q}_{j}^k(t')\mathcal{Q}_{j}^{k,\dagger} (t)}$, one can then obtain a generalized energy transfer rate
\begin{align}
\label{App.B1.Eq1}
[\kappa_{\text{ET}}]^{jj'}=\sum_{\{m_k=0\}}^\infty\sum_{\{l_k=0\}}^\infty \prod_{k=1}^n e^{-2\lambda_k^2}\frac{\lambda_k^{2(m_k+l_k)}}{m_k!l_k!}\frac{\sum_{k=1}^n(m_k+l_k)\Gamma_k\Omega_{jj'}^2}{[\sum_{k=1}^n (m_k+l_k)\Gamma_k]^2+[\delta_j-\delta_{j'}-\sum_{k=1}^n(m_k+l_k)\nu_k]^2},
\end{align}
where the sums go over all indices $\{m_k\}=m_1,\hdots,m_n$ and $\{l_k\}=l_1,\hdots,l_n$.


\section{Equations of motion for $\mathcal{N}$ molecules. Vacuum-Rabi splitting at high densities.}
\label{B2}
We derive the equations of motion only for individual operators ($a$ and $\sigma_j$) for any excitation level and justify the single excitation approximation. Also we consider individual decay of the emitters and low excitation, i.e., $\sigma^{z}_j \approx -1$.\\
From the master equation we obtain the equations of motion for an operator $O$ via
\begin{subequations}
\begin{align}
\label{App.D.Eq1}
\dot{\braket{O}} = \text{Tr}\left[\dot{\rho} O \right]
= \text{Tr}\left[\left(\frac{i}{\hbar}[\rho,H] + \mathcal{L}_{\text{c}}[\rho] + \mathcal{L}_{\text{e}}[\rho] + \mathcal{L}_{\text{FRET}}[\rho] \right) O \right],
\end{align}
\end{subequations}
where $\mathcal{L}_{\text{FRET}}[\rho] = \sum_{i \neq j} \kappa_{\text{ET}}^{ij}\left(2 \sigma^{\dagger}_{j}\sigma_{i}\rho \sigma^{\dagger}_{i}\sigma_{j} - \sigma^{\dagger}_{i}\sigma_{i}\sigma_{j}\sigma^{\dagger}_{j} \rho - \rho \sigma^{\dagger}_{i}\sigma_{i}\sigma_{j}\sigma^{\dagger}_{j} \right)$.\\
For $O = \sigma_{j}$ we derive
\begin{subequations}
\begin{align}
\label{App.D.Eq2}
\text{Tr}\left[\mathcal{L}_{\text{FRET}}[\rho] \sigma_{j} \right] &= -\sum_{k\neq j} \kappa_{\text{ET}}^{jk}\braket{\sigma_{j}} + \sum_{k \neq j}\left(\kappa_{\text{ET}}^{jk} - \kappa_{\text{ET}}^{kj} \right)\braket{\sigma_j \sigma^{\dagger}_k \sigma_k},
\end{align}
\end{subequations}
while for $O = \sigma^{\dagger}_j \sigma_j$ we obtain
\begin{subequations}
\begin{align}
\label{App.D.Eq3}
\text{Tr}\left[\mathcal{L}_{\text{FRET}}[\rho] \sigma^{\dagger}_j \sigma_{j} \right] &= -\sum_{k\neq j} 2\kappa_{\text{ET}}^{jk} \braket{\sigma^{\dagger}_j\sigma_{j}} + \sum_{k\neq j} 2\kappa_{\text{ET}}^{kj} \braket{\sigma^{\dagger}_k\sigma_{k}} + \sum_{k \neq i,j}2\left(\kappa_{\text{ET}}^{jk} - \kappa_{\text{ET}}^{kj} \right)\braket{\sigma^{\dagger}_j \sigma_j \sigma^{\dagger}_k \sigma_k}.
\end{align}
\end{subequations}
This results in the equations of motion
\begin{subequations}
\begin{align}
\label{App.D.Eq4}
\dot{\braket{a}} &= -(\kappa + i \delta)\braket{a} -i \sum_{j}g^{*}_{j}\braket{\sigma_{j}} + \eta \\
\dot{\braket{\sigma_j}} &= -\left( \gamma + \sum_{k\neq j} \kappa_{\text{ET}}^{jk} +i(\delta_l+\delta_j) \right)\braket{\sigma_j} - ig_j \braket{a} + \sum_{k \neq j}\left(\kappa_{\text{ET}}^{jk} - \kappa_{\text{ET}}^{kj} \right)\braket{\sigma_j \sigma^{\dagger}_k \sigma_k},
\end{align}
\end{subequations}
where $\delta = \omega_{l} - \omega_{c}$ and $\delta_l = \omega_{l} - \omega_{e}$. Additionally we obtain for the population
\begin{subequations}
\begin{align}
\label{App.D.Eq5}
\dot{\braket{\sigma^{\dagger}_j \sigma_j}} &= -2\left(\gamma + \sum_{k \neq j}\kappa_{\text{ET}}^{jk} \right)\braket{\sigma^{\dagger}_j \sigma_j} + \sum_{k\neq j} 2\kappa_{\text{ET}}^{kj} \braket{\sigma^{\dagger}_k\sigma_{k}} + i\left(g^{*}_{j}\braket{a^{\dagger}\sigma_{j}} - g_{j} \braket{a \sigma^{\dagger}_{j}} \right) +
\sum_{k \neq j} 2\left[\kappa_{\text{ET}}^{jk} - \kappa_{\text{ET}}^{kj} \right]\braket{\sigma^{\dagger}_{k}\sigma_{k}\sigma^{\dagger}_{j}\sigma_{j}},
\end{align}
\end{subequations}
which in general for small population approximates to
\begin{equation}
\dot{P}_j \approx -2\left(\gamma + \sum_{k \neq j}\kappa_{\text{ET}}^{jk} \right)P_j + \sum_{k\neq j} 2\kappa_{\text{ET}}^{kj}P_k + i\left(g^{*}_{j}\braket{a^{\dagger}\sigma_{j}} - g_{j} \braket{a \sigma^{\dagger}_{j}} \right),
\end{equation}
where we have defined $P_j = \braket{\sigma^{\dagger}_j \sigma_j}$.\\
These results agree with the expressions obtained by starting from the general quantum Langevin equations where we obtain
\begin{subequations}
\begin{align}
\label{App.D.Eq6}
\dot{a} &= -(\kappa + i \delta)a -i \sum_{j}g^{*}_{j}\sigma_{j} + \eta + \sqrt{2\kappa} a_{\text{in}} \\
\dot{\sigma_j} & = -\left( \gamma + \sum_{k\neq j} \kappa_{\text{ET}}^{jk} + i(\delta_l +\delta_j) \right)\sigma_j - ig_j a + \sum_{k \neq j}\left(\kappa_{\text{ET}}^{jk} - \kappa_{\text{ET}}^{kj} \right) \sigma^{\dagger}_k \sigma_k \sigma_j -\sqrt{2\gamma}\sigma^{\text{in}}_j \\\nonumber
&+ \sum_{k \neq j} \left[\sqrt{2\kappa_{\text{ET}}^{jk}}\xi^{\dagger}_{\text{in}, jk}\sigma_{k} - \sqrt{2\kappa_{\text{ET}}^{kj}}\sigma_{k} \xi_{\text{in},kj} \right],
\end{align}
\end{subequations}
where $a_{\text{in}}$, $\sigma_{\text{in}}$ and $\xi_{\text{in},ij}$ are the noise operators for the collapse operators $a$, $\sigma$ and $\sigma^{\dagger}_i\sigma_j$, respectively.\\
For low population, we can linearize the equations of motion for the molecule-cavity system
\begin{subequations}
\begin{align}
\label{App.D.Eq6}
\dot{\alpha} &\approx -(\kappa + i \delta)\alpha -i \sum_{j}g^{*}_{j}\beta_{j} + \eta \\
\dot{\beta_j} & \approx -\left( \gamma + \sum_{k\neq j} \kappa_{\text{ET}}^{jk} + i(\delta_l +\delta_j) \right)\beta_j - ig_j \alpha,
\end{align}
\end{subequations}
which results in the cavity transmission
\begin{equation}
t = \kappa\left[\kappa + i\delta + \sum^{\mathcal{N}}_{j=1}\frac{|g_j|^2}{\gamma + \sum_{k\neq j}\kappa^{jk}_{\text{ET}} + i(\delta_l + \delta_j)} \right]^{-1}.
\end{equation}
Additionally, we obtain for the equations of motion in the bright dark basis the expressions
\begin{subequations}
\begin{align}
\dot{\mathcal{D}}_k &= -i\left(\delta_l + \bar{\delta} -i\left(\gamma + \frac{1}{\mathcal{N}}\sum^{\mathcal{N}}_{j=1}\sum^{\mathcal{N}}_{j' \neq j}\kappa^{jj'}_{\text{ET}}\right) \right)\mathcal{D}_k -i\sum^{\mathcal{N}-1}_{k'\neq k} \tilde{\Delta}_{kk'}\mathcal{D}_{k'} - i\tilde{\Delta}_{k\mathcal{N}}\mathcal{B}, \\
\dot{\mathcal{B}} &= -i\left(\delta_l + \bar{\delta} -i\left(\gamma + \frac{1}{\mathcal{N}}\sum^{\mathcal{N}}_{j=1}\sum^{\mathcal{N}}_{j' \neq j}\kappa^{jj'}_{\text{ET}}\right) \right)\mathcal{B}-i\sum^{\mathcal{N}-1}_{k'=1} \tilde{\Delta}_{\mathcal{N}k'}\mathcal{D}_{k'} - i\sqrt{\mathcal{N}}g\alpha, \\
\dot{\alpha} &= -i(\delta -i\kappa)\alpha -i\sqrt{\mathcal{N}}g^{*}\mathcal{B} + \eta,
\end{align}
\end{subequations}
where $\tilde{\Delta}_{kk'} = (1/\mathcal{N})\sum^{\mathcal{N}}_{j=1}\left(\delta_j - i\sum_{j'\neq j} \kappa_{\text{ET}}^{jj'}\right) e^{-i2\pi j(k-k')/\mathcal{N}}$. By performing the same steps as introduced in previous chapters where we have injected the solution for the dark modes at steady state into the bright mode, we finally obtain
\begin{subequations}
\begin{align}
\label{App.D.Eq7a}
\dot{\mathcal{B}} &= -i\left(\delta_l + \bar{\delta} -i\left(\gamma + \bar{\kappa}_{\text{ET}}\right) - \sum^{\mathcal{N}-1}_{k,k'=1}\tilde{\Delta}_{\mathcal{N}k}\tilde{\mathcal{M}}^{-1}_{kk'}\tilde{\Delta}_{k'\mathcal{N}} \right)\mathcal{B}- i\sqrt{\mathcal{N}}g\alpha, \\
\label{App.D.Eq7b}
\dot{\alpha} &= -i(\delta -i\kappa)\alpha -i\sqrt{\mathcal{N}}g^{*}\mathcal{B} + \eta,
\end{align}
\end{subequations}
where $\tilde{\mathcal{M}}_{kk'} = (\delta_l - i\gamma)\delta_{kk'} + \tilde{\Delta}_{kk'}$ for $k,k' \in \{1,\dots, \mathcal{N}-1\}$, $\bar{\kappa}_{\text{ET}} = \frac{1}{\mathcal{N}}\sum^{\mathcal{N}}_{j=1}\sum^{\mathcal{N}}_{j' \neq j}\kappa^{jj'}_{\text{ET}}$ and $\delta_{\text{dark}} + i\gamma_{\text{dark}} = \sum^{\mathcal{N}-1}_{k,k'=1}\tilde{\Delta}_{\mathcal{N}k}\tilde{\mathcal{M}}^{-1}_{kk'}\tilde{\Delta}_{k'\mathcal{N}}$. This allows us to obtain the vacuum Rabi splitting (VRS) in the case of disorder and FRET transfer by diagonalizing the corresponding matrix to Eq.~\eqref{App.D.Eq7a} and Eq.~\eqref{App.D.Eq7b} and setting $\delta_l = \delta = 0$, which is given by
\begin{equation}
\label{App.D.Eq7c}
\text{VRS} = \Im\left\{2\sqrt{\frac{\left((\gamma + \bar{\kappa}_{\text{ET}} + \gamma_{\text{dark}}-\kappa) + i(\bar{\delta} -\delta_{\text{dark}}) \right)^2}{4}-\mathcal{N}|g|^2} \right\}.
\end{equation}\\
We can get qualitative expressions for the FRET induced rates from the following procedure. In the case of two molecules with only one vibrational mode each we can use the expression for the energy transfer rate in Eq.~\eqref{App.B1.Eq1} to obtain
\begin{eqnarray}
\label{App.D.Eq8a}
\sum_{jj'} \kappa^{jj'}_{\text{ET}} &=& \sum_{n_1}^\infty\sum_{n_2}^\infty s^{\lambda}_{n_1}s^{\lambda}_{n_2}(n_1 + n_2)\Gamma\sum^{\infty}_{j' \neq j}\frac{\Omega_{jj'}^2}{[(n_1+n_2)\Gamma]^2+[\delta_j-\delta_{j'}-(n_1+n_2)\nu]^2}.
\end{eqnarray}
Considering a homogeneous distribution of molecules, in the term for the dipole-dipole interaction $\Omega_{jj'} = (3/2)\gamma/(k|\mathbf{r}_j - \mathbf{r}_{j'}|)^3$ we can set $\mathbf{r}_j$ to zero, since each molecule witnesses the same surrounding environment. Additionally, by using the probability distribution $p(r,\delta)$ to find a molecule at a distance $r$ with detuning $\delta$ we exchange the sum over $j'$ with the integration
\begin{equation}
\frac{1}{\mathcal{N}} \sum_{j'\neq j} \approx \int^{R}_{r_{\text{min}}}dr \int^{\infty}_{-\infty} d\delta p(r,\delta),
\end{equation}
where $R$ is the radius of a spherical volume $V$ and $r_{\text{min}}$ is the minimal radius that follows from the volume $V/\mathcal{N}$ that each molecule occupies individually.
Assuming that $r$ and $\delta$ are independent we can rewrite $p(r,\delta) = p(r)p(\delta)$, where $p(r) = 4\pi r^2/V$ and $p(\delta) = (1/\sqrt{2\pi w^2})e^{-(\delta/\sqrt{2w^2})^2}$. This allows us to obtain
\begin{eqnarray}
\nonumber
\sum_{jj'} \kappa^{jj'}_{\text{ET}} &\approx& \sum_{n_1}^\infty\sum_{n_2}^\infty s^{\lambda}_{n_1}s^{\lambda}_{n_2}(n_1 + n_2)\Gamma \mathcal{N} \int^{R}_{r_{\text{min}}}dr p(r) \left(\frac{3\gamma}{2(kr)^3}\right)^2\int^{\infty}_{-\infty}d\delta  \frac{p(\delta)}{[(n_1+n_2)\Gamma]^2+[\delta_j-\delta_{j'}-(n_1+n_2)\nu]^2}.\\
&\approx& \mathcal{N}^2 \pi\left(\frac{3\gamma}{2(kR)^3}\right)^2 \sum_{n_1}^\infty\sum_{n_2}^\infty s^{\lambda}_{n_1}s^{\lambda}_{n_2}V(\delta_j -(n_1+n_2)\nu ; w, (n_1+n_2)\Gamma),
\end{eqnarray}
where $V(x;w, \gamma)$ describes a Voigt profile. We find here that $\sum_{jj'} \kappa^{jj'}_{\text{ET}}$ is proportional to $\mathcal{N}^2$ which allows us to find a qualitative expression for the VRS in Eq.~\eqref{App.D.Eq7b} that can be used for fitting.

\end{document}